\documentclass[preprint,showpacs,preprintnumbers,amsmath,amssymb,
floatfix]{revtex4}

\usepackage{graphicx}
\usepackage{dcolumn}
\usepackage{bm}
\usepackage{slashed}
\def\Id{{\rm 1\kern-.3em I}}

\begin{document}

\title{Final-state interactions in semi-inclusive deep inelastic scattering off
the Deuteron}

\author{W. Cosyn}
\email{Wim.Cosyn@UGent.be}
\altaffiliation{On leave from: Department of Physics and Astronomy,
 Ghent University, Proeftuinstraat 86, B-9000 Gent, Belgium}
\author{M. Sargsian}

\affiliation{Department of Physics, Florida International University, Miami,
Florida 33199, USA}
\date{\today}

\begin{abstract}
Semi-inclusive deep inelastic scattering off the Deuteron  with production of 
a slow nucleon in recoil kinematics is studied in 
the virtual nucleon approximation, in which 
the final state interaction (FSI)  
is calculated within generalized eikonal approximation.
The cross section is derived in a factorized
approach, with a factor describing the virtual photon interaction with the
off-shell nucleon and a distorted spectral function accounting for the
final-state interactions.  
One of the main goals of the study is to understand how much the general
features of 
the diffractive high energy soft rescattering accounts for the observed 
features of FSI in deep inelastic scattering(DIS). 

Comparison with  the Jefferson Lab data shows good
agreement in the covered range of kinematics. Most importantly, our calculation 
correctly reproduces the rise of the FSI in the forward direction of 
the slow  nucleon production angle.

By fitting our calculation to the data we  extracted the 
$W$ and $Q^2$ dependences 
of  the total cross section and slope factor of the interaction of 
DIS products, $X$,   off the spectator nucleon. This analysis shows 
the $XN$ scattering cross section rising with $W$ and decreasing with an
increase of 
$Q^2$. 
Finally, our analysis points at a largely suppressed 
off-shell part of the rescattering amplitude. 

\end{abstract}


\maketitle 

\section{Introduction} 
\label{sec:intro}

In recent years, a process that has garnered a fair amount of attention is the
$d(e,e'p_s)X$ reaction at high $Q^2$.  In this reaction, deep inelastic
scattering (DIS) occurs on a constituent of the deuteron and a slow
\emph{spectator} proton is detected in coincidence with the scattered electron. 
This reaction can be used in several ways to study the role of the QCD dynamics
at
nucleonic length scales.  At very small spectator proton momenta, the
DIS occurs on a nearly on-shell neutron and it allows one to extract information
about the \emph{``free''} neutron structure function $F_{2N}$ in a way that
minimizes
the nuclear effects inherent to a bound neutron.  Detailed information about the
neutron structure function helps to constrain the QCD models of the nucleon 
and can be used to
determine the relative $d$ to $u$ quark densities at large Bjorken $x$.  At
larger
spectator momenta, high density configurations of the deuteron will occur in
which the proton and neutron are in very close proximity to each other.  Under
these circumstances the partonic structure of nucleons could strongly
modify \cite{Melnitchouk:1996vp} 
with the  possibility of two nucleons merging into six quark configurations 
at asymptotically large relative momenta in the deuteron
\cite{Carlson:1994ga,Carlson:1999uk}. 
Consequently, experiments
that explore these kinematics can be used to study the modifications of nucleon
properties and the role of quark degrees of freedom in these situations.  Two
recent Jefferson Lab Hall B experiments have studied the $d(e,e'p_s)X$
reactions: one at high
\cite{Klimenko:2005zz}, and the other  at low spectator momenta
\cite{Bonus:2003}.  New measurements will be possible after the 12 GeV
upgrade of JLab is completed.

In experiments exploring the partonic structure of 
the nucleon, one generally wants to have
kinematics that minimize the final-state interactions (FSI) of the produced
$X$-states
with the spectator  nucleon as this FSI make the extraction of the observable
one is
looking for less straightforward.  On the other hand the $d(e,e'p_s)X$ reaction
in 
kinematics that favor larger contributions from FSI can be used in order to
study
the process of hadronization.  The attenuation of the  produced hadronic state 
by the spectator when compared to the free process 
can yield   information on the space-time structure of the hadronization
process.  
Thus in this respect FSI becomes very important part of the semi-inclusive DIS 
process.
To quantify the effects of
FSI in DIS, model calculations are needed and this has
already resulted in the development of several theoretical approaches 
\cite{Simula:1996xk,Melnitchouk:1996vp,Sargsian:2005rm,CiofidegliAtti:1999kp,
CiofidegliAtti:2002as,CiofidegliAtti:2003pb,Palli:2009it,Atti:2010yf}.

The major problem one faces in calculations of FSI of DIS products with the
spectator 
nucleon in $d(e,e^\prime N)X$  reactions is the lack of the detailed
understanding 
of the composition and space-time evolution  of  the  hadronic system produced 
after the deep inelastic scattering of the virtual photon off the  bound
nucleon. Moreover
both the composition and space-time evolution are function of the Bjorken $x$
and $Q^2$
probed in the reaction.

In this paper  we study the question on  how much the final state interaction of
the 
DIS products are defined by the general properties of soft reinteractions. In
other words, 
how far we can go with the description of FSI without knowing the specific
properties of 
the hadronic intermediate state after the  initial DIS scattering?
Based on the general properties of the reaction
a  factorized approach is used in the calculations, whereby the cross section is
split into the parts describing the interaction of the virtual photon with a
bound nucleon and the distorted spectral function which includes the effect of
 final-state interactions. The deep inelastic interaction with moving bound
nucleon 
is calculated  within the virtual nucleon approximation while the FSI are
included using 
the framework of generalized
eikonal approximation (GEA)
\cite{Frankfurt:1994kt,Frankfurt:1996xx,Sargsian:2001ax}.  

The paper is organized as follows.  In Sec.~\ref{sec:formalism} we describe
the general properties of 
the reaction and main assumptions based on which we derive the plane-wave
impulse 
approximation and final-state 
interaction parts of the scattering.  An overview of the various approximations
used in this derivation is also given.  In
Sec. \ref{sec:results}, the results of our model calculations are
discussed and compared to the data from the \emph{Deeps} experiment performed at
JLab
\cite{Klimenko:2005zz}.  Finally, conclusions are given in
Sec. \ref{sec:conclusion}.

\section{formalism}
\label{sec:formalism}

\subsection{General Structure  of the Reaction}
We consider the process
\begin{equation}\label{eq:process}
 e+d\rightarrow e'+p_s+X, 
\end{equation}
in which incoming electron $e$ has energy $E_e$, while $E_{e'}$ and $\theta_e$
denote the energy and scattering angle of the final electron $e'$. We define
the lab frame
four-momenta of the involved particles as $p_D\equiv(M_D,0)$ for the deuteron, $
q\equiv (\nu,\vec{q})$ for the virtual photon (with the z-axis chosen along
$\vec{q}$),
$p_s\equiv(E_s=\sqrt{\vec{p}_s^2+m_p^2},\vec{p}_s)$ for
the spectator proton and
$p_x\equiv(E_X,\vec{p}_X)=(\nu+M_D-E_s,\vec{q}-\vec{p}_s)$ the center of mass
momentum
of the undetected produced hadronic system $X$.  We can express the 
differential cross section
for process (\ref{eq:process}) through the four independent DIS structure
functions 
in the following form:
\begin{multline} \label{eq:cross}
 \frac{d\sigma}{dxdQ^2d\phi_{e^\prime}\frac{d^3p_s}{2E_s(2\pi)^3}}=\frac{
2\alpha_{\text{EM}}^2}{xQ^4}
(1-y-\frac{x^2y^2m_n^2}{Q^2})\left(F^D_L(x,Q^2)+(\frac{Q^2}{2|q|^2}+
\tan^2{\frac{\theta_e}{2}})\frac{\nu}{m_n}F^D_T(x , Q^2)+\right.\\
\left.\sqrt { \frac { Q^2 } { |q
|^2}+\tan^2{\frac{\theta_e}{2}} } \cos{\phi}
F^D_{TL}(x,Q^2)+\cos{2\phi}F^D_{TT}(x,Q^2)\right)\,.
\end{multline}
Here, $\alpha_{\text{EM}}$ is the fine-structure constant,
$-Q^2=\nu^2-\vec{q}^2$ is the
four-momentum transfer, Bjorken $x=\frac{Q^2}{2m_nx}$ (with $m_n$ the mass of
the neutron), $y=\frac{\nu}{E_e}$, and $\phi$ is the angle between the
scattering
~$(e,q)$
 and reaction~$(q,p_s)$ planes .  

We now define the nuclear electromagnetic tensor as
\begin{multline}\label{eq:htensor}
  W^{\mu\nu}_D=\frac{1}{4\pi M_D}\frac{1}{3}\sum_X\sum_{s_s,s_x,s_D}\langle
Ds_D|J^{\dagger
\mu}|X s_x,p_s s_s\rangle \langle X s_x,p_s s_s | J^\nu|Ds_D\rangle\\
\times(2\pi)^4\delta^4(q+p_D-p_s-p_x) d^3\tau_x\,,
\end{multline}
with $d^3\tau_x$ a phase-space factor for $X$, and $s_D,s_s$, and $s_x$ the
spin projection of the deuteron, spectator proton and $X$ respectively.  
The four deuteron semi-inclusive structure functions
$F^D_i(x,Q^2)$
are related to components
of the nuclear electromagnetic tensor $W_D^{\mu\nu}$ as follows:
\begin{align}
F^D_L(x,Q^2)&=\nu\frac{Q^4}{|q|^4}W_D^{00}(x,Q^2)\,,\nonumber\\
F^D_T(x,Q^2)&=m_n(W_D^{xx}(x,Q^2)+W_D^{yy}(x,Q^2))\,,\nonumber\\
F^D_{TL}(x,Q^2)\cos\phi&=-2\nu\frac{Q^2}{|q|^2}W_D^{0x}(x,Q^2)\,,\nonumber\\
F^D_{TT}(x,Q^2)\cos2\phi&=\nu\frac{Q^2}{2|q|^2}(W_D^{xx}(x,Q^2)-W_D^{yy}(x,
Q^2))\, .\label
{eq:F_D}
\end{align}

\subsection{Main Approximations}
\label{subsec:approx}

In the further derivations  we  use the  following approximations which are
based largely on the 
general properties of DIS scattering  as well as properties  
of the subsequent small angle rescattering of the 
fast moving hadronic  system  off the  slow recoil nucleon:

\begin{itemize}
\item  [-]{\em virtual nucleon approximation:} To treat the electromagnetic 
interaction with the bound nucleon in the deuteron we use the virtual nucleon 
approximation~(VNA) in which it is assumed that the virtual photon interacts
with
the off-shell  nucleon in the deuteron while the second nucleon is on its 
mass shell \cite{Melnitchouk:1996vp,Sargsian:2005rm,Sargsian:2001gu}.  
The VNA is based on the following main assumptions: (i) only the $pn$ component
of the
deuteron wave function is considered in the reaction, 
(ii) the negative energy projection of the virtual
nucleon propagator gives negligible contribution to the scattering amplitude,
and (iii) interactions of the virtual photon with exchanged mesons is 
neglected. Assumptions (i) and (ii) can be satisfied when the momentum of the
spectator proton is limited to $p_s \leq 700$ MeV/c \cite{Sargsian:2009hf}, 
while (iii) is satisfied at large $Q^2$ ($> 1$ GeV$^2$)
\cite{Sargsian:2001ax,Sargsian:2002wc}.

The electromagnetic tensor  of the $\gamma N$ interaction is off-shell and 
the gauge  invariance is restored by expressing the longitudinal component of
the 
electromagnetic current through its  $0$'th component as follows:
\begin{equation}
J^3 = {\frac{q_0}{q_3}}J^0\,.
\label{eq:gauge}
\end{equation}
The nuclear  wave  function in the VNA is  normalized to account for the baryon
number 
conservation \cite{Frankfurt1981215,Frankfurt:1976gz,Frankfurt1987254,
Landshoff:1977pg}:
\begin{equation}
\int \alpha |\Psi_D(p)|^2 d^3p  = 1,
\label{norm}
\end{equation}
where $\alpha= 2-\frac{2(E_s-p_{s,z})}{M_D}$ is the light cone momentum
fraction of the deuteron 
carried by the bound nucleon normalized in such a way that the half of the
deuteron 
momentum fraction corresponds to $\alpha=1$.
Because of the virtuality of interacting nucleon it is impossible 
to satisfy the momentum  sum rule at the same time. As a result
\begin{equation}
\int \alpha^2 |\Psi_D(p)|^2 d^3p  < 1,
\end{equation}
which can be qualitatively  interpreted as part  of the deuteron momentum
fraction being 
distributed to non-nucleonic degrees of freedom which are unaccounted for within
the VNA.

\begin{figure}[htb]
\begin{center}
  \includegraphics[width=0.5\textwidth]{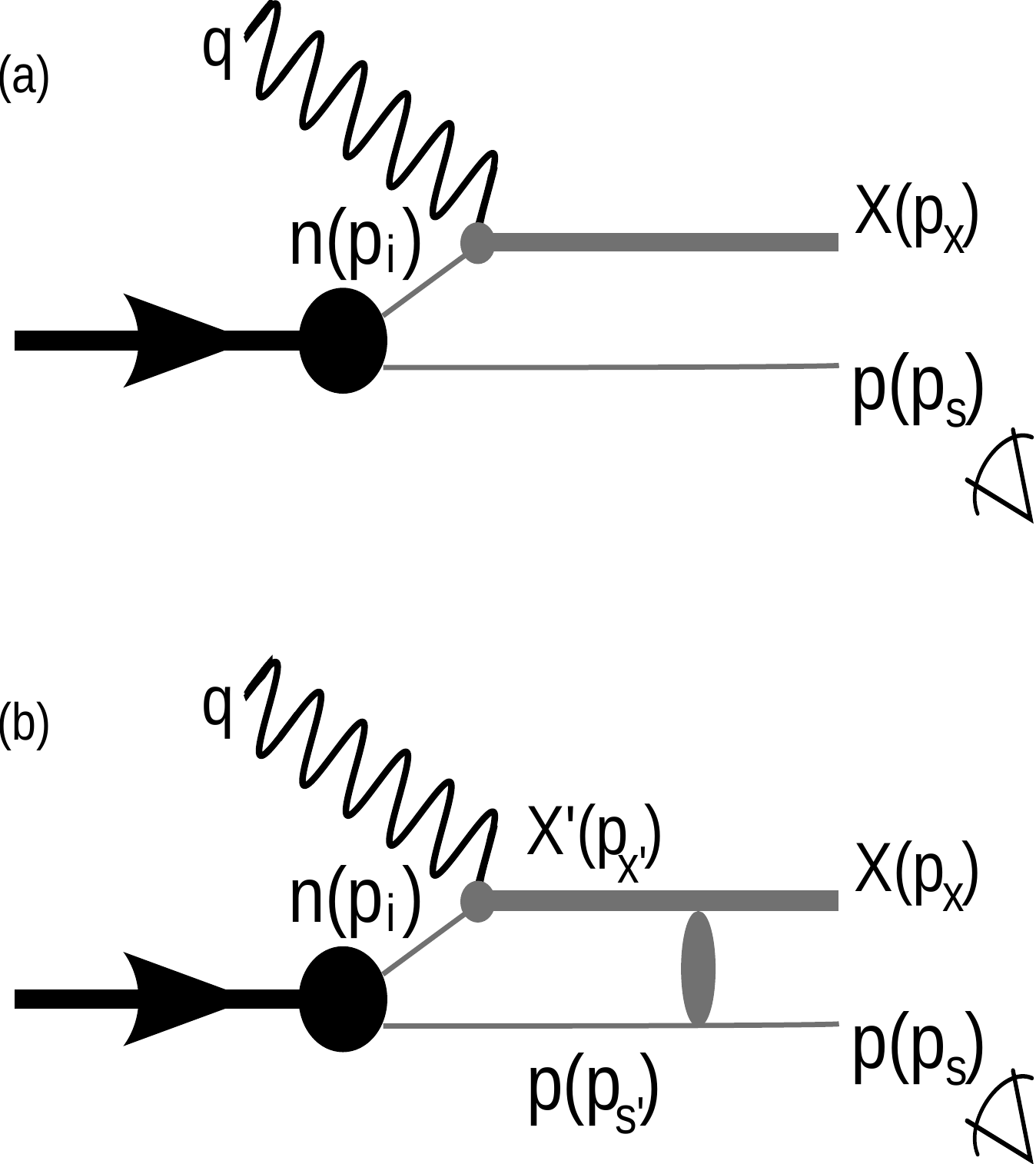}
\caption{Diagrams entering in the model for the $d(e,e'p_s)X$ reaction. Panel
(a) shows the plane-wave contribution.  Panel (b) shows the FSI term.}
\label{fig:diagrams}       
\end{center}
\end{figure}

By applying the VNA for calculation of  the matrix element $\langle X s_x,p_s
s_s | J^\mu|Ds_D\rangle$,
we can limit the Feynman diagrams taken into account to
those of Fig.~\ref{fig:diagrams}  in which 
\ref{fig:diagrams}(a) represents the plane-wave impulse approximation (PWIA)
diagram. Here DIS occurs on the neutron and the proton is left in the on-shell
positive energy state without further interaction in the final state.  
The diagram of Fig.~\ref{fig:diagrams}(b) shows again DIS on the
neutron, which is afterwards followed by a $X'p \rightarrow Xp$ rescattering.  
In calculating this diagram we have to sum over the all possible intermediate 
$X^\prime $ states.

The calculation of the final-state interactions is based on the following main
assumptions for the rescattering diagram of Fig.~\ref{fig:diagrams}(b).
\item [-] {\em diffractive form of the rescattering amplitude:} In the
considered reaction 
the FSI represents a  small angle rescattering of the DIS products off the slow
spectator nucleon.
It is in principle a very complex problem to account for  the details of the
interaction of the intermediate 
``$X^\prime$'' state since its structure depends on the $Q^2$ ($x_{Bj}$) and the
produced mass $W$ of the 
$\gamma ^* N$ reaction.  However in the limit  where the produced intermediate
and final masses are small compared 
to the transferred momenta:
\begin{equation}
q\gg M_{X^\prime},  M_{X^\prime}, 
\label{qgMM}
\end{equation}
one can assume that the  propagation of 
the produced hadronic system is eikonal and the general structure of the small
angle rescattering is 
diffractive.  
The approximation of Eq.(\ref{qgMM})  allows one to model the FSI amplitude of
the hadronic
$X^\prime$ system in the following  form:
\begin{equation}\label{eq:scatter}
\sum\limits_{X^\prime}f_{X^\prime N, X N} =  f_{XN}(t,Q^2,x_{Bj}) =
\sigma_{\text{tot}}(Q^2,x_{Bj})(i + \epsilon(Q^2,
x_{Bj}))e^{\frac{B(Q^2,x_{Bj})}{2} t},
\end{equation}
where  the sum of the all possible  $X^\prime N\rightarrow X N$ amplitudes 
are represented in the  effective diffractive amplitude form,
$f_{XN}(t,Q^2,x_{Bj}) $
with   effective total cross section
$\sigma_{\text{tot}}$,  real part, $\epsilon$ and slope factor $B$. 
A similar 
approximation is used for 
the FSI studies in semi-inclusive DIS scattering
\cite{Simula:1996xk,Melnitchouk:1996vp,Sargsian:2005rm,CiofidegliAtti:1999kp,
CiofidegliAtti:2002as,CiofidegliAtti:2003pb,Palli:2009it} as well as for
studies 
of color transparency phenomena in 
which the  intermediate state represents an off-shell  coherent composite
system 
with reduced interaction cross
section (see e.g. \cite{Farrar:1988me,Frankfurt:1994hf,
Frankfurt:1994kt,Frankfurt:1997ss,Cosyn:2007er,Ryckebusch:2007zz,
Gallmeister:2010wn}).  In principle, a more elaborate model which sums the
contribution of different resonances as e.g. in Ref.~\cite{Frankfurt:1992zp}
could be used but this would go beyond the goal in this paper of describing the
reaction with the basic elements of high-energy rescattering.

\item [-]{\em factorization:} In the situation in which momentum transfer in
DIS exceeds the 
 momentum of the recoil slow nucleon one can factorize DIS scattering from the
amplitude 
of the  final state interaction. Such an approximation commonly referred as
distorted 
wave impulse approximation~(DWIA) is  valid in the limit of $\sqrt{Q^2}\gg p_s$
in which case the
electromagnetic current is insensitive to the momentum of the stuck nucleon. 
The validity of  the DWIA was checked quantitatively for quasielastic scattering
in the case of 
$d(e,e'N)N$ reactions \cite{Jeschonnek:2000nh,Sargsian:2009hf}. These
calculations 
demonstrated that for $Q^2=2-4$~GeV$^2$ factorization approximation works
reasonably well for up to $p_s=400$~MeV/c and then at larger momenta it  
systematically underestimates the FSI contribution as compared to 
the prediction based on an unfactorized calculation.
The underestimation can be understood qualitatively, since in the 
case of nonfactorization the amplitude of electromagnetic interaction 
enters in the FSI amplitude at smaller values of bound nucleon momenta 
and therefore predicts more rescattering than the DWIA does. This pattern 
one also expects to be generally valid for inelastic interactions.

\item [-]{\em approximate conservation law of high energy small angle
scatterings:} In the eikonal regime of  small 
angle scattering there is an  approximate conservation law for  the ``$-$'' 
component
\footnote{The \emph{$\pm$} components of the 
momentum is defined as $p_\pm = E \pm p_{z}$.}
of slow nucleon momenta  involved in the scattering \cite{Sargsian:2001ax}.
According to this law, because 
the fast particle  attains its momentum after the small-angle scattering the
slow nucleon will conserve its ``$-$`` component. 
This follows from  the conservation of the ``-'' component of the total
momentum 
in $X^\prime N^\prime \rightarrow  XN$ scattering and relations  
$p_{X^\prime-}  \approx {\frac{m_{X^\prime}^2 +p_{X^\prime\perp}^2}{2q}} \ll 1$ 
and $p_{X^-}  \approx   {\frac{{m_{X}^2 + p_{X\perp}^2}}{2q}} \ll 1$ 
provided that the condition of Eq.~(\ref{qgMM}) is satisfied.
This yields:
\begin{equation}
p_{s^\prime-} - p_{s-} = p_{X^-} - p_{X^\prime-} \approx 0\,.
\label{mcomp}
\end{equation}
Using this  relation and assuming that 
\begin{equation}
p^2_{s\perp} < k^2_\perp
\label{psperp}
\end{equation}
where $k^2_\perp$ is the average transferred  momentum in the rescattering 
one obtains:
\begin{equation}
m^2_{X} = (p_{X^\prime} + p_{s^\prime} - p_{s})^2  \approx m^2_{X^\prime} -
2p_{X^\prime\perp}(p_{s^\prime\perp}-p_{s\perp}) - k^2_\perp 
\approx m^2_{X^\prime} + k^2_\perp > m^2_{X^\prime}
\label{eq:masses}
\end{equation}
where in the above derivation we used the fact that in the limit of
Eq.~(\ref{psperp}) 
$p_{X^\prime\perp} = - p_{s^\prime\perp} \approx k_\perp$.  The above result
qualitatively means that in the
situation in which two collinear particles are produced  by the  diffractive
scattering of a fast and slow particle 
with equal  and opposite transverse momenta the  mass of the final fast particle
is larger than the initial mass.

Using the characteristic  values of the diffractive slope, $B = 4-6$~GeV$^{-2}$,
one can estimate $k_{\perp,RMS}\approx 500-600$~MeV/c.  This estimate of
$k_\perp$ and 
Eq.~(\ref{psperp})  further constrains the values of spectator nucleon momenta
for which 
the  calculations will be valid.
\end{itemize}

Our derivations in the following two subsections are based on the above
assumptions.

\subsection{Plane-wave impulse approximation}
\label{subsec:pwia}

Applying Feynman diagram rules (see e.g. Ref.~\cite{Sargsian:2001ax}) and
introducing the 
effective wave functions of the final hadronic system $X$, the amplitude of the
PWIA 
diagram in Fig.~\ref{fig:diagrams}(a) takes the following form:
\begin{equation}\label{eq:PWIA}
 \langle X s_x,p_s s_s |
J^\mu|Ds_D\rangle^{\text{PWIA}}=-\bar{\Psi}_X(p_X,s_X)\Gamma^\mu_{\gamma^*X}
\frac {
\slashed{p}_i+m_n } {p^2_i-m_n^2}\cdot \bar{u}(p_s,s_s)\Gamma_{DNN}\cdot
\chi^{s_D}\,.
\end{equation}
Here, $\Psi_X$ is a wave function for $X$ and $\Gamma^\mu_{\gamma^*X}$
represents the
electromagnetic vertex of the DIS.  The transition of the deuteron into a
$pn$ system is described by the vertex function $\Gamma_{DNN}$ and $\chi^{s_D}$
denotes the spin wave function of the deuteron.  The lab frame four-momentum of
the
struck neutron $p_i$ in the PWIA is defined as
\begin{equation}
 p_i=(M_D-E_s,-\vec{p_s})\,.
\end{equation}

We now split the initial nucleon propagator in on-shell and off-shell parts
by adding and subtracting an on-shell energy part:
\begin{equation}\label{eq:redux}
 \slashed{p}_i+m_n=
\slashed{p}_i^{\text{on}}+m_n+(E_i^{\text{off}}-E_i^{\text{on}})\gamma^0\,,
\end{equation}
with $E_i^{\text{off}}=M_D-E_s$ and $E_i^{\text{on}}=\sqrt{m_n^2+p_s^2}$.  Next
we write
\begin{align}\label{eq:redux2}
 \slashed{p}_i^{\text{on}}+m_n&=\sum_{s_i}u(p_i,s_i)\bar{u}(p_i,s_i)\,,\\
(E_i^{\text{off}}-E_i^{\text{on}})\gamma^0&\approx\frac{E_i^{\text{off}}-E_i^{
\text{on}}}{2m_n}\gamma^0\sum_{s_i}u(p_i,s_i)\bar{u}(p_i,s_i)\,,
\end{align}
where in the last equation we used $\sum_{s_i}u(p_i,s_i)\bar{u}(p_i,s_i)
\approx 2m_nI$, which is consistent with neglecting the negative energy
component 
of the  bound nucleon propagator.  Now, with the
definition \cite{Gribov:1968gs,Bertocchi:1972}
\begin{equation}\label{eq:deuterondef}
 \Psi_D^{s_D}(p_1s_1,p_2s_2)=-\frac{\bar{u}(p_1,s_1)\bar{u}(p_2,s_2)\Gamma_{
DNN}\cdot
\chi^{s_D}}{(p_1^2-m_1^2)\sqrt{2}\sqrt{2(2\pi)^3(p_2^2+m_2^2)^{\frac{1}{2}}}}\,,
\end{equation}
we can write Eq.~(\ref{eq:PWIA}) as
\begin{multline}\label{eq:PWIAredux}
 \langle X s_x,p_s s_s |
J^\mu|Ds_D\rangle^{\text{PWIA}}= \sqrt{2}
\sqrt{(2\pi)^3 2E_s}\sum_{s_i}\langle X
s_x|\Gamma^\mu_{\gamma^* N,X}|p_i s_i \rangle
\left(1+\frac{E_i^{\text{off}}-E_i^{\text{on}}}{2m_n}
\gamma^0\right)\\\times\Psi^{s_D}
_D(p_i s_i, p_s
s_s)\,.
\end{multline}
Even though Eq.~(\ref{eq:PWIA}) is gauge invariant, one can not calculate the
off-shell part of the current explicitly
since the form  of the electromagnetic vertex $\Gamma^\mu_{\gamma^*N,X}$ is
unknown. 
Instead, in the reminder of the derivation the term ${\frac{E^{off}_i -
E^{on}_i}{2m_n}}\gamma_0$ 
associated with the   off-shell behavior of the photon-neutron interaction will
be dropped and the 
gauge invariance will be restored through Eq.~(\ref{eq:gauge}). Inserting
Eq.~(\ref{eq:PWIAredux}) in Eq.~(\ref{eq:htensor}) we obtain for
the PWIA contribution
\begin{multline}\label{eq:htensor2}
  W^{\mu\nu}_D=\frac{1}{4\pi M_D}\frac{2}{3}\sum_{X}\sum_{s_s,s_x,s_D,s_i,s'_i}
\langle p_i s_i |\Gamma^{\dagger\mu}_{\gamma^* N,X}|X
s_x\rangle \langle X
s_x|\Gamma^\nu_{\gamma^* N,X}|p_i s'_i \rangle\\
\times(2\pi)^4\delta^4(q+p_i-p_x) d^3\tau_x(2\pi)^3 2E_s \Psi^{\dagger s_D}(p_i
s_i, p_s
m_s) \Psi^{ s_D}(p_i
s'_i, p_s
m_s)\,.
\end{multline}
We can simplify this further by using 
\begin{equation}
\sum_{s_D,s_s} \Psi^{\dagger s_D}(p_i
s_i, p_s
m_s) \Psi^{ s_D}(p_i
s'_i, p_s
m_s)=\sum_{s_D,s_s}\left|\Psi^{ s_D}(p_i
s_i, p_s
s_s)\right|^2 \delta_{s_i,s'_i}
\end{equation}
 and 
\begin{equation}\sum_{s_D,s_s}\left|\Psi^{ s_D}(p_i
s_i=+1, p_s
s_s)\right|^2 =  \sum_{s_D,s_s}\left|\Psi^{ s_D}(p_i
s_i=-1, p_s
s_s)\right|^2\,.
\end{equation}
Eq.~(\ref{eq:htensor2}) then becomes
\begin{multline}\label{eq:htensor3}
  W^{\mu\nu}_D=\frac{1}{4\pi M_D}\sum_X\sum_{s_x,s_i}
\langle p_i s_i |\Gamma^{\dagger\mu}_{\gamma^* N,X}|X
s_x\rangle \langle X
s_x|\Gamma^\nu_{\gamma^* N,X}|p_i s_i \rangle(2\pi)^4\delta^4(q+p_i-p_x)
d^3\tau_x\\
\times(2\pi)^3 2E_s \frac{1}{3}\sum_{s_D,s_s,s'_i}\left|\Psi^{s_D}(p_i
s'_i, p_s
s_s)\right|^2\,.
\end{multline}
After
defining the spectral function
\begin{equation}
 S(p_s)\equiv \frac{1}{3}\sum_{s_D,s_s,s_i}\mid
\Psi^{ s_D}(p_i
s_i, p_s
s_s)\mid^2\,,
\end{equation}
and using the following expression for the nuclear tensor of the DIS process
on a moving nucleon
\begin{equation} \label{eq:fact1}
  W^{\mu\nu}_N=\frac{1}{4\pi m_n}\frac{1}{2}\sum_X\sum_{s_x,s_i}\langle p_i
s_i|\Gamma^{\dagger \mu}_{\gamma^*N,X}|X s_x \rangle \langle X
s_x|\Gamma^{\nu}_{\gamma^*N,X}|p_i s_i \rangle (2\pi)^4 \delta^4(q+p_i-p_x)
d^3\tau_X\,.
\end{equation}
we can write Eq.~(\ref{eq:htensor3}) as
\begin{equation}\label{eq:fac}
 W^{\mu\nu}_D=W^{\mu\nu}_N S(p_s)(2\pi)^32E_s\,,
\end{equation}
where we also used $\frac{2m_n}{M_D}\approx1$.  Substituting Eq.~(\ref{eq:fac})
in~(\ref{eq:F_D}) allows us to relate the four deuteron
DIS structure functions to the nucleon structure functions. After
straightforward
calculations one obtains the following relations for the deuteron structure
functions:
\begin{align}
 F_L^D(x,Q^2)&= \left[ (\alpha_i+\frac{\alpha_q(p_i\cdot
q)}{Q^2})^2(1+\cos{\delta})^2\frac{\nu}{\hat{\nu}}F_{2N}(\alpha_i,\hat{x},
Q^2)-\frac{\nu}
{m_n} \sin^2 { \delta}F_{1N}(\alpha_i,\hat{x},Q^2)\right]\nonumber\\
&\times S(p_s)(2\pi)^32E_s\,,\\
F_T^D(x,Q^2)&=
\left(2F_{1N}(\alpha_i,\hat{x},Q^2)+\frac{p_T^2}{m_n\hat{\nu}}F_{2N}(\alpha_i,
\hat{x},
Q^2)\right)S(p_s)(2\pi)^32E_s\,,\\
F_{TT}^D(x,Q^2)&=
\frac{\nu}{\hat{\nu}}\frac{p_T^2}{m_n^2}\frac{\sin^2\delta}{2}F_{2N}(\alpha_i,
\hat{x},
Q^2)S(p_s)(2\pi)^32E_s\,,\\
F_{TL}^D(x,Q^2)&=2(1+\cos\delta)\frac{p_T}{m_n}(\alpha_i+\frac{
\alpha_q(p_i\cdot q)} {Q^2 }
)\frac{\nu}{\hat{\nu}}F_{2N}(\alpha_i,\hat{x},Q^2)
S(p_s)(2\pi)^32E_s\,,
\end{align}
where $\alpha_i = \frac{2p_i^-}{M_D}$, $\alpha_q =
\frac{2q^-}{M_D}$, $\hat{\nu}=\frac{p_i\cdot q}{m_n}$,
$\hat{x}=\frac{Q^2}{2m_n\hat{\nu}}$,$\cos\delta =
\frac{\nu}{|q|}$, $\sin^2\delta =\frac{Q^2}{|q|^2}$, and $F_{1N},F_{2N}$ are
the effective nucleon structure functions, which are defined at $\hat{x}$ and
in principle could be modified due to the nuclear binding (see e.g.
Ref.~\cite{Melnitchouk:1996vp}).

Note that the inclusive $F_2$, $F_1$ and $F^{in}_L$ structure functions  of the 
deuteron can be obtained from the above given semi-inclusive structure 
functions through the following relations:
\begin{eqnarray}
F_{2,D} & = & \sum\limits_{N}\int \left[F_L^D + {\frac{Q^2} {2
|q|^2}}{\frac{\nu}
{m_N}}F_T^D\right]
{\frac{d^3p_s} {(2\pi)^2 2E_s}} \approx \sum\limits_{N}\int \left[F_L^D +
xF_T^D\right]
{\frac{d^3p_s} {(2\pi)^2 2E_s}}, \nonumber  \\
F_{1,D} & = & \sum\limits_{N}\int {\frac{F_T^D}{2}} {\frac{d^3p_s} {(2\pi)^2
2E_s}}
\nonumber \\ 
F^{in}_{L,D} & \equiv & F_{2,D}-2xF_{1,D} = 
\sum\limits_{N}\int \left[F_L^D + ({\frac{Q^2}{2 |q|^2}}{\frac{\nu}
{m_N}}-x)F_T^D\right]
{\frac{d^3p_s} {(2\pi)^2 2E_s}} \nonumber \\ &\approx& \sum\limits_{N}\int F_L^D
{\frac{d^3p_s} {(2\pi)^2 2E_s}},
\end{eqnarray}
where one sums by the contributions of both the proton and neutron. The L.H.S.
parts of 
the equations represent the expressions in the case of the Bjorken limit with
$x$ fixed and
$Q^2,\nu \rightarrow \infty$.

\subsection{Final-state interaction amplitude}
\label{subsec:fsi}
With the same notations as in Sec. \ref{subsec:pwia}, we can write for the
amplitude of the FSI diagram in Fig.~\ref{fig:diagrams}(b)
\begin{multline}\label{eq:FSI}
  \langle X s_x,p_s s_s |
J^\mu|Ds_D\rangle^{\text{FSI}}=-\sum_{X^\prime}\int
\frac{d^4p_{s'}}{i(2\pi)^4}\frac{\bar{\Psi}_X(p_X,s_X)\bar{u}(p_s,s_s)F_{X'N,XN
}[\slashed{p}_{s'}+m_p]}
{[p^2_{s'}-m_p^2+i\epsilon]}\\
\times
\frac{
G(P_{X'})
\Gamma^\mu_{\gamma^*X'}[\slashed
{p}_{i'}+m_n] \Gamma_{DNN}\cdot
\chi^{s_D} }{[p^2_{X'}-m_{X'}^2+i\epsilon][p^2_{i'}-m_n^2+i\epsilon]}\, ,
\end{multline}
where $G(p_{X'})$ describes the Green's function of the intermediate state $X'$ 
which has four-vector $p_{X'}\equiv
p_{i'}+q = (\nu+M_D-E_{s'},\vec{q}-\vec{p}_{s'})$ and a mass $m_{X^\prime}$,
while the
intermediate struck neutron has four-vector
$p_{i'}=(M_D-E_{s'},-\vec{p}_{s'})$. $F_{X'N,XN
}$ represents the invariant $X'N\rightarrow XN$ scattering amplitude which is 
expressed in the following form
\begin{align}
 F_{X'N,XN}(s,t)&=\sqrt{(s-(m_n-m_{X'} )^2)(s-(m_n+m_ { X' } )^2)
} f_ { X'N,XN}(s,t)\nonumber
\\
&=\beta(s,m_{X'})f_{X'N,XN}(s,t),,
\end{align}
with $s=(p_X+p_s)^2=(p_{X'}+p_{s'})^2$ the total invariant energy of the
scattering system and the scattering amplitude $f_{X'N,XN}$ defined such that
$\text{Im}\left[f_{X'N,XN}(t\equiv 0)\right]=\sigma_{\text{tot}}$, where
$\sigma_{\text{tot}}$ represents the total cross section of the scattering of
the produced $X'$ system off the spectator nucleon. Based on the assumptions
of the VNA from Sec. \ref{subsec:approx}, the
intermediate spectator nucleon can be placed on the nucleon mass-shell by
integrating $d^0p_{s'}$ through the positive energy pole only:
\begin{equation}
 \int \frac{d^0p_{s'}}{p^2_{s'}-m_p^2+i\epsilon}\rightarrow
-i\frac{\pi}{E_{s'}}\,.
\end{equation}
This allows us to use the on-shell spinor relation
$\slashed{p}_{s'}+m_p=\sum_{s_{s'}}u(p_{s'},s_{s'})\bar{u}(p_{s'},s_{s'})$ in
the nominator of Eq.~(\ref{eq:FSI}).  For the propagator of the initial neutron
we again use the prescription of Eqs.~(\ref{eq:redux}) to (\ref{eq:redux2}). 
For the intermediate state $X'$ an on-shell relation for the Green's function 
$G(p_{X'})=\sum_{s_{x'}}\psi(p_{x'},s_{x'})\psi^\dagger(p_{x'},s_{x'})$
is used as in the high $Q^2$ limit the off-shell contribution in
Eq.~(\ref{eq:FSI}) becomes small due to the large momentum involved in the 
propagator of the intermediate state $X^\prime$.  By making use of
$p_X^2=(q+p_D-p_s)^2=m_X^2$, the
denominator of the $X'$ propagator can be rewritten as
\begin{equation}
 p_{X'}^2-m_{X'}^2+i\epsilon=2\mid\vec{q}\mid(p_{s',z}-p_{s,z}
+\Delta+i\epsilon)\,,
\end{equation}
with
\begin{equation}
 \Delta=\frac{\nu+M_D}{\mid\vec{q}\mid}(E_s-E_{s'})+\frac{m_{X}^2-m_{X'}^2}{
2\mid\vec { q } \mid }\,.
\label{Delta}
\end{equation}

 All this combined with
the deuteron wave function of Eq.~(\ref{eq:deuterondef}) allows us to write the
FSI amplitude as
\begin{multline}
 \langle X s_x,p_s s_s |
J^\mu|Ds_D\rangle^{\text{FSI}}=-\sum_{X^\prime}\sum_{s_{i},s_{s'},s_{x'}}
\int
\frac{d^3p_{s'}}{(2\pi)^3}\beta(s,m_{X'})\langle X
s_x, p_s s_s |
f_{ X'N,XN}(s,t)| X' s_{x'}, p_{s'} s_{s'} \rangle \\\times
\frac{\langle X' s_{x'}|\Gamma^\mu_{\gamma^* N,X'}|p_{i'} s_{i}
\rangle\Psi^{s_D}(p_{i'} s_{i}, p_{s'} s_{s'})}
{4E_{s'}\mid\vec{q}\mid[p_{s',z}-p_{s,z}
+\Delta+i\epsilon]}
\sqrt{2}\sqrt{(2\pi)^3 2E_{s'}}\,.
\end{multline}
In a next step, we assume the rescattering amplitude conserves the helicities of
all particles involved
\begin{equation}
 \langle X
s_x, p_s s_s |
f_{ X'N,XN}(s,t)| X' s_{x'}, p_{s'} s_{s'} \rangle \approx \langle X
s_x, p_s s_s |
f_{ X'N,XN}(s,t)| X' s_{x}, p_{s'} s_{s} \rangle \delta_{s_s,s_{s'}}
\delta_{s_x,s_{x'}}\,,
\end{equation}
and we use the following approximation to take the current matrix element out
of the integration:
\begin{equation}
 \langle X' s_{x}|\Gamma^\mu_{\gamma^* N,X'}|p_{i'} s_{i}\rangle \approx 
 \langle X s_{x}|\Gamma^\mu_{\gamma^* N,X}|p_{i} s_{i}\rangle\,.
\end{equation}
This allows us to factorize the nuclear tensor again like in
Eq.~(\ref{eq:fac}). For the sum of the plane-wave and FSI amplitudes, we then
obtain
\begin{equation}
  W^{\mu\nu}_D=W^{\mu\nu}_N S^{\text{dist.}}(p_s)(2\pi)^32E_s\,,
\end{equation}
with the distorted spectral function defined as
\begin{multline}\label{eq:distspectral}
  S(p_s)^{\text{dist.}}\equiv \frac{1}{3}\sum_{s_D,s_s,s_i} \left|
\Psi^{s_D}(p_i s_i, p_s s_s) -
\sum_{X^\prime}\int\frac{d^3p_{s'}}{(2\pi)^3}\frac{\beta(s,m_{X'})}{4\mid\vec{q}
\mid\sqrt{E_sE_
{ s' } } }\right.\\\left.\times \langle X
s_x, p_s s_s |
f_{ X'N,XN}(s,t)| X' s_{x}, p_{s'} s_{s}\rangle \frac{\Psi^{s_D}(p_{i'} s_{i},
p_{s'} s_{s})}{[p_{s',z}-p_{s,z}
+\Delta+i\epsilon]}\right|^2\,.
\end{multline}

\subsection{Distorted spectral function}
\label{subsec:dist}
For calculation of the distorted spectral function in Eq.(\ref{eq:distspectral})
we use VNA model of 
deuteron wave function of Eq.~(\ref{eq:deuterondef}) which 
can be  related to the
non-relativistic deuteron wave function by
\cite{Frankfurt:1976gz,Sargsian:2005rm,Sargsian:2009hf}
\begin{equation}
 \Psi_D(p)= \Psi_D^{\text{NR}}(p) \sqrt{\frac{M_D}{2(M_D-E_s)}}\,,
\end{equation}
which explicitly conserves the baryonic sum rule of Eq.~(\ref{norm}).
The parameterizations for the non-relativistic wave function used in this paper
all take the following form (see e.g.
Refs.~\cite{Machleidt:2000ge,Lacombe:1980dr}):
\begin{equation}
  \Psi_D^{s_D}(\vec{p}s_1,-\vec{p}s_2) =
\chi^{\dagger,s_1}\chi^{\dagger,s_2}\left[\sum_j \frac{c_j}{p^2+m_j^2}
 + \sum_j\frac{d_j}{p^2+m_j^2}
\mathcal{S}(\vec{p})\right] \chi^{s_D}\,,
\end{equation}
with $\mathcal{S}(\vec{p})=\sqrt{\frac{1}{8}}\left(
\frac{3\vec{\sigma}_1\cdot\vec{p}\vec{\sigma}_2\cdot\vec{p}}{p^2}-\vec{\sigma}
_1\cdot\vec{\sigma}_2 \right)$ the tensor operator.  
Such form allows to perform the $dp_{s',z}$ integration
in the distorted spectral function of Eq.~(\ref{eq:distspectral}) 
analytically by making use of the pole structure of these parameterizations as
well as  
the pole of the propagator in Eq.~(\ref{eq:distspectral})
at $\tilde{p}_{s',z}=p_{s,z}-\Delta$.  
In the latter case the mass of the produced intermediate state $m_{X^\prime}$
enters in the 
phase factor $\Delta$. We note that  even though we sum over the all possible
intermediate states $X^\prime$ the mass $m_{X^\prime}$ is  defined by the four
momenta of the interacting virtual nucleon and virtual photon $q$. 
Based on the assumption that FSI is dominated by a small angle diffractive
scattering,  
the  phase factor $\Delta$ 
is evaluated based on the property of the approximate conservation law of
``$-$'' components of  rescattering particle momenta discussed in
Sec.\ref{subsec:approx}. 
Taking into account the 
relation of Eq.~(\ref{eq:masses}) in the definition of the $\Delta$ factor   in 
Eq.~(\ref{Delta}) we evaluate:
\begin{align}
 \Delta&=\frac{\nu+M_D}{\mid\vec{q}\mid}(E_s-m_p)+\frac{m_X^2-\gamma}{
2\mid\vec { q } \mid } &\text{for}\, \gamma \leq m_X^2\,,\nonumber\\
\Delta&=\frac{\nu+M_D}{\mid\vec{q}\mid}(E_s-m_p) &\text{for}\, \gamma >
m_X^2\,,
\label{deltas}
\end{align}
where $\gamma\equiv m_{X'}^2(p_{i'}=0)= m_n^2+2m_n\nu-Q^2$ is the produced DIS
mass  off the stationary nucleon.  
The latter approximation for $m_{X^\prime}$ is justified by the fact that due to
the peaking of 
the deuteron wave function at small momenta the integrand in
Eq. (\ref{eq:distspectral}) is 
dominated  by smaller virtual nucleon momenta   than in the PWIA term

The $p_{s^\prime,z}$ integration in Eq. (\ref{eq:distspectral}) is performed
analytically by 
closing the integration contour into either the upper or lower complex
hemispheres.
In both cases \footnote{See Appendix B of Ref. \cite{Sargsian:2001ax} for
details
of integration.} one 
obtains:
\begin{equation}
\int dp_{s',z}
\frac{\Psi^{s_D}(p_{i'}s_i,p_{s'}s_s)}{p_{s',z}-p_{s,z}+\Delta+i\epsilon}
= -i\pi \Psi^{s_D}(\tilde{p}_{s,z},p_{s',\perp},s_i,s_s) - \pi \tilde{p}_{s,z}
\tilde{\Psi}(\tilde{p}_{s,z},p_{s',\perp},s_i,s_s)
\label{pszint}
\end{equation}
where the distorted wave function  $\tilde{\Psi}$ is defined in Eq
(\ref{psidist}). 
In the above  equation the first term can be identified with the (imaginary)
on-shell part of the 
$p_{s',z}$ propagator while the second term with the (real) principal value
integration.
Inserting Eq. (\ref{pszint}) into Eq. (\ref{eq:distspectral})  one obtains for
the
distorted spectral
function:
\begin{multline}
  S(p_s)^{\text{dist.}}= \frac{1}{3}\sum_{s_D,s_s,s_i} \left| \Psi^{s_D}(
 \vec{p}_s,s_i,
s_s) +\frac{i}{2}
\sum_{X^\prime}\int\frac{d^2p_{s',\perp}}{(2\pi)^2}\frac{\beta(s,m_{X'})}{
4\mid\vec{q}\mid\sqrt
{ E_sE_
{ s' } } }\right.\\\left.\times\left[ \langle X, p_s |
f^{\text{on}}_{X'N,XN}(s,t)| X', \tilde{p}_{s'}\rangle
\Psi^{s_D}(\tilde{p}_{s'},
s_{i}, 
s_{s}) \right.\right.\\ \left.\left. -i \langle X, p_s |
f^{\text{off}}_{ X'N,XN}(s,t)| X' , \tilde{p}_{s'} \rangle \tilde{p}_{s',z}
\tilde{\Psi}^{s_D}(\tilde{p}_{s'},s_{i},
s_{s})\right]\right|^2\,,
\label{distS}
\end{multline}
where for the distorted wave function of the deuteron one obtains:
\begin{equation}
 \tilde{\Psi}^{s_D}(p_s,s_1,
s_2)= \left[u_1(p_s)+\frac{w_1(p_s)}{\sqrt{8}}
\mathcal{S}(p_s)
+\frac{w_2(p_s)}{\sqrt{8}}\frac{p^2_{s,\perp}}{p^2_{s,z}}\left(\mathcal{S}
(p_s)-\mathcal { S } (p_ { s , \perp } )\right)\right] \chi^{s_1}\chi^{s_2} \,,
\label{psidist}
\end{equation}
with
\begin{align}
 u_1(p)&=\sum_j\frac{c_j}{\sqrt{p^2_\perp+m_j^2}p^2+m_j^2}\,,\nonumber\\
w_1(p)&=\sum_j\frac{d_j}{\sqrt{p^2_\perp+m_j^2}p^2+m_j^2}\,,\nonumber\\
w_2(p)&=\sum_j\frac{d_j}{m_j^2\sqrt{p^2_\perp+m_j^2}}\,.
\end{align}

The distorted spectral function in  Eq.(\ref{distS}) depends on the intermediate
state 
$X^\prime$ through the final state rescattering amplitude only. As a result one
can factorize 
the  sum over the $X^\prime$ in the form of Eq.(\ref{eq:scatter}) and 
represent 
the on-shell forward scattering amplitude in the form of 
\begin{equation}
f^{on}_{XN} = \sigma_{\text{tot}}(Q^2,W)(i +
\epsilon(Q^2,W))e^{\frac{B(Q^2,W)}{2} t}\,,
\label{fon}
\end{equation}
with $W$ the invariant mass of the produced hadronic state $X$.
 For the off-shell amplitude
$f^{\text{off}}$ there is
no clear prescription, but  following our main goal of studying the
semi-inclusive DIS 
based only on basic properties of the high-energy scattering we identify two
extreme cases 
for off-shell part of the rescattering amplitude, one when it is taken to be
zero 
({\em no off-shell FSI}) and the other in which off-shell amplitude is assumed
to be equal to 
the on-shell amplitude $f^{\text{on}}_{XN}$ referred as {\em maximal off-shell
FSI}.
The numerical estimates for $f^{\text{off}}_{XN}$ we used in
 our calculations will be discussed below in Sec.~\ref{sec:results}.

With this, we have all the ingredients needed to compute the cross section of
Eq.~(\ref{eq:cross}).

\section{Results}
\label{sec:results}

\subsection{Experimental observables}

In this section, we compare calculations in our model with the first results
extracted
from data taken in the \emph{Deeps} experiment at JLab \cite{Klimenko:2005zz}. 
Events
of the data set were binned in $Q^2$, $p_s$, $\cos \theta_s$ (with $\theta_s=
\widehat{q,p_s}$) and
$\hat{x}$ (or the invariant mass of the produced hadronic state $W$). 
In order to compare our model calculations
with
the data, we integrate Eq.~(\ref{eq:cross}) over $\phi_{e'}$, use
\begin{equation}
 \frac{d\hat{x}}{dx}=\frac{2\hat{x}^2}{x}\frac{\nu}{|q|}\left|
\frac{\alpha_i}{\alpha_q}+\frac{1}{2\hat{x}}\right|\,, 
\end{equation}
and relate $F_{1N}(\alpha_i,\hat{x},Q^2)$ to $F_{2N}(\alpha_i,\hat{x},Q^2)$ for
a moving nucleon:
\begin{equation}
 F_{1N}(\alpha_i,\hat{x},Q^2)=\frac{2\hat{x}}{1+R}\left[\left(\frac{\alpha_i}{
\alpha_q}+\frac{1}{2\hat{x}} \right)^2-\frac{p_T^2}{2Q^2}R
\right]F_{2N}(\alpha_i,\hat{x},Q^2)\,,
\end{equation}
where $R=\frac{\sigma_L}{\sigma_T}\approx 0.18$ is the ratio of the
longitudinal to transverse cross sections for scattering off the nucleon.
This yields for the differential cross section:
\begin{multline} \label{eq:crossdeeps}
 \frac{d\sigma}{d\hat{x}dQ^2d^3p_s}=\frac{4\pi\alpha_{\text{EM}}}{Q^4\hat{x}}
\frac{|q|}{m_n}
\left(1-y-\frac { x^2y^2m_n^2} {Q^2}
\right)\left(\frac{Q^2}{|q|^2}+\frac{2\tan^2\
\frac{\theta_e}{2}}{1+R} \right)\left|\frac{\alpha_i}{
\alpha_q}+\frac{1}{2\hat{x}} \right|^{-1}\\\times\left[\left(\frac{\alpha_i}{
\alpha_q}+\frac{1}{2\hat{x}} \right)^2+\frac{p_T^2}{2Q^2}
\right]F_{2N}(\alpha_i,\hat{x},Q^2)S(p_s)\,.
\end{multline}
Now using  Eq.~(\ref{eq:crossdeeps}) we need to reproduce 
the quantity $F_{2N}P(\vec{p}_s)$ (with
$P(\vec{p}_s)=\frac{\alpha_iM_D}{2(M_D-E_S)}|\Psi_D^{\text{NR}}(p_s)|^2$) 
for which the experimental data are given in Ref.~\cite{Klimenko:2005zz}. For
this  
we divide the cross section of Eq.~(\ref{eq:crossdeeps}) with the following
prefactor
\begin{equation}
 \mathcal{F}=\frac{4\pi\alpha_{\text{EM}}}{Q^4\hat{x}}\left[\frac{\hat{y}}{
2(1+R)}+(1-\hat{y
})+\frac {
p_i^2\hat{x}^2\hat{y}^2 } {Q^2 }\frac{1-R}{1+R} \right]\,,
\end{equation}
where $\hat{y}=\frac{p_i\cdot q}{p_i\cdot k_e}$.  This results to the 
following representations of our model calculations:
\begin{equation}
 (F_{2N}P)_{\text{model}}=\frac{1}{\mathcal{F}}
\left(\frac{d\sigma}{d\hat{x}dQ^2d^3p_s}\right)_{\text{model}}\,.
\label{exp}
\end{equation}

In numerical estimates we use the SLAC parameterizations for the neutron
structure functions $F_{1N}$
and $F_{2N}$ \cite{PhysRevD.20.1471} in the calculations as these were used in
the analysis of the \emph{Deeps} data \cite{Klimenko:2005zz}.  The arguments of
the
nucleon structure
functions are defined from the off-shell kinematics $(p_i+q)=W^2$, where the
four momentum of the initial nucleon is defined as $p_i=p_D-p_s$.  No
additional modifications due to nuclear modifications like the EMC effect are
assumed for the nucleon structure functions. This is in accordance to our 
approach of estimating the properties of the reaction based on the basic 
properties of the high energy scattering rather than modeling the specific
details 
of the reaction. Additionally  we deem the influence of these
modifications small in comparison with the typical magnitude of the
experimental uncertainties to extract unambiguous information here.

\subsection{Numerical Estimates}

\begin{figure}[ht]
\begin{center}
\includegraphics[width=0.7\textwidth]{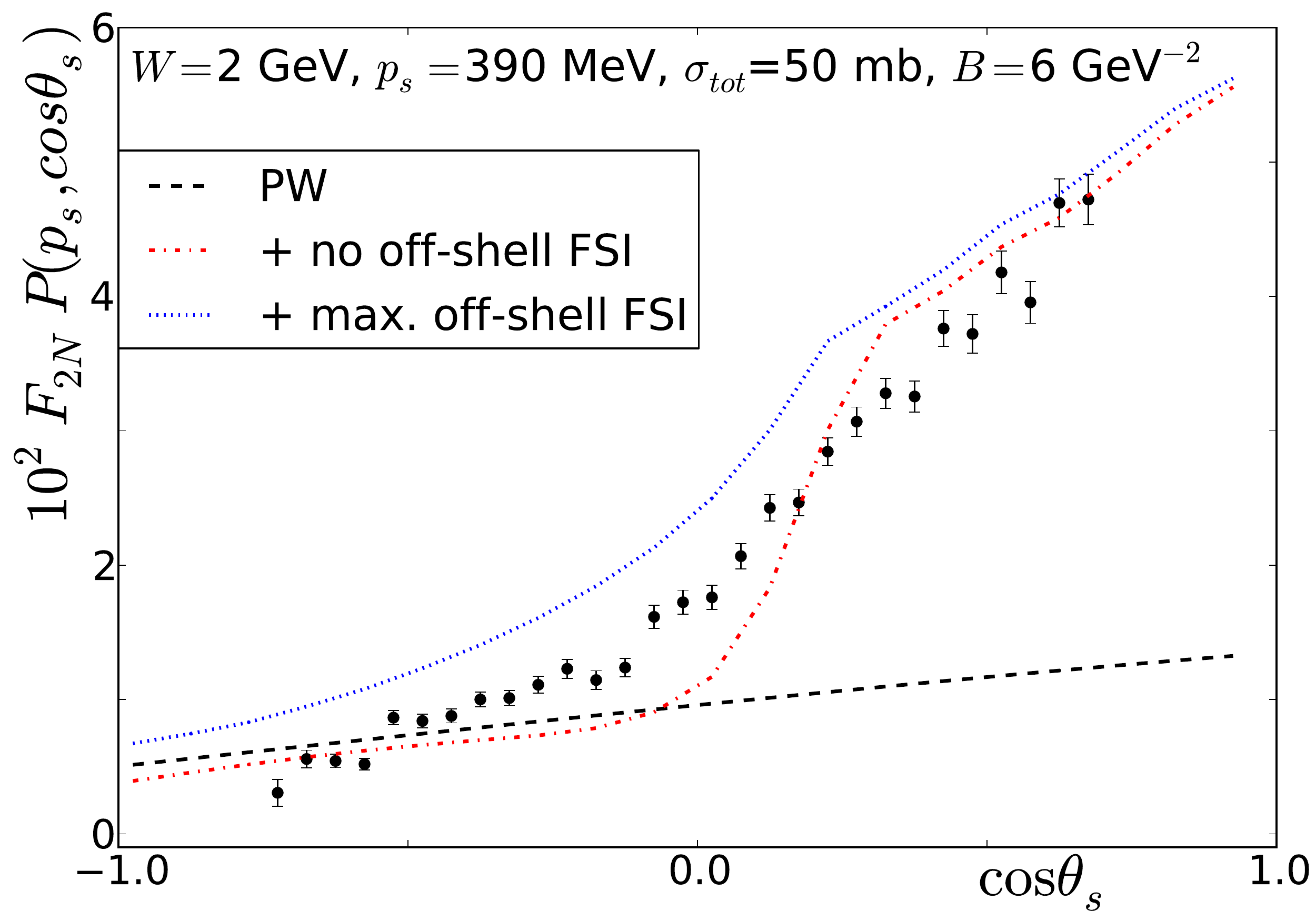}
\caption{(Color online) Comparison between the \emph{Deeps} data
\cite{Klimenko:2005zz} and model calculations at $Q^2=1.8~\text{GeV}^2$.  The
dashed black curve is a
plane-wave calculation, the other include final-state interactions.  The
effective total cross section and slope parameter in the final-state
interaction amplitude are fixed to $\sigma_{\text{tot}}=50mb$, $B=6
\text{GeV}^{-2}$ and $\epsilon = -0.5$.  The dotted blue curve has an off-shell
rescattering amplitude amplitude equal to the on-shell one (\emph{maximal
off-shell FSI}), the dash-dotted red curve has \emph{no off-shell FSI}.}
\label{fig:typical}       
\end{center}
\end{figure}

We start first with the calculation of the quantity of Eq.~(\ref{exp}) for
the typical kinematic 
setting of the experiment \cite{Klimenko:2005zz} with $Q^2=1.8$~GeV$^2$, $W^2=
2$~GeV and $p_s=390$~MeV/c.
For on-shell part of the rescattering $XN\rightarrow XN$ amplitude  we use the
diffractive form 
of the parameterization of Eq.~(\ref{eq:scatter}) with characteristic  values
of 
$\sigma_{tot} = 50$~mb, $B = 6$~GeV$^-2$ and $\epsilon = -0.5$. Our estimate of
the total $XN$ cross section 
is based on the assumption that the final state consists of the hadronic state
equivalent to 
one nucleon and one pion. In principle it is possible to develop specific model 
(see e.g. Ref.~\cite{Atti:2010yf}) describing the $XN$ rescattering, however we
follow 
here our main goal of understanding how far we can go with describing data on
basic properties of 
high energy scattering.  For the off-shell part of the $XN$ rescattering we use
two limiting cases
as discussed above: $f_{XN}^{off-shell}=0$  ({\em no off-shell FSI }) and 
$f_{XN}^{off-shell}=f_{XN}^{on-shell}$ ({\em maximal off-shell FSI}).  The
results of these calculations 
are given in Fig.~\ref{fig:typical}. 

As the figure shows,   FSI effects  continuously grow in the forward angles of
production of
recoil proton.   This result is strikingly different  from the case of 
the quasielastic $d(e,e'N)X$ scattering in which
case the FSI is  maximal at transverse angles ($\sim 70^0$) 
of recoil nucleon production (see e.g. \cite{Sargsian:2009hf}) and
diminishes in  the forward direction.   
The continuously growing FSI contribution in the forward direction for 
DIS scattering follows from  the specific structure of the  phase factors 
($\Delta$) entering in Eq.~(\ref{deltas}) which follows from
Eq.~(\ref{eq:masses}).  
For forward angles the dominant mass contribution $m_X'$ decreases, as can be
seen in the stationary 
approximation (virtual photon energy $\nu$ decreases with forward angles). 
As $m_X'$ decreases the off-diagonal mass term in Eq. (\ref{deltas}) grows
bigger and so does $\Delta$ causing the 
peak to  shift to more forward angles.

Another observation from  Fig.~\ref{fig:typical} is the relatively small
contribution 
due to the off-shell part of the $XN$ rescattering amplitude. This result is in
agreement 
of the space-time analysis of high energy small angle rescattering of
Ref.~\cite{Frankfurt:2008zv}, 
according to which the longitudinal distances 
that off-shell particle propagates before rescattering significantly
shrinks in the high $Q^2$ and fixed Bjorken $x$ limit. This results in 
the suppression of the off-shell part of the FSI.

\medskip
\medskip
\medskip

The next question we address is whether the parameters of $X'N$ rescattering 
amplitude are sensitive to the produced DIS mass $W$ and $Q^2$.
For this  we assume some of the parameters entering the rescattering
amplitude of Eq.~(\ref{eq:scatter}) to be free parameters.  We made fits using
one (effective total cross section $\sigma_{\text{tot}}$) or two
($\sigma_{\text{tot}}$ and slope factor
$B$) free parameters.  The real to imaginary part ratio of the amplitude was
fixed at $\epsilon=-0.5$, a value extrapolated from nucleon-nucleon scattering
parameterizations.  

For the off-shell rescattering amplitude in addition to above mentioned 
\emph{no off-shell FSI} and \emph{maximal off-shell FSI} options we consider
the 
third approach in which case we 
parameterize the off-shell amplitude as
\begin{equation} \label{eq:offrescatter}
 f_{X N}^{\text{off}} =
\sigma_{\text{tot}}^{\text{on}}(Q^2,W_N)(i +
\epsilon^{\text{on}}(Q^2,W_N))e^{-\frac{B^{\text{off}}(Q^2,W_N)}{2} t}\,,
\end{equation}
were the effective cross section and real part parameters were taken equal to
the on-shell ones, but the slope parameter was taken as a new free parameter in
the fit.  This will give us some measure of the size of the suppression
as compared to the on-shell amplitude, this approach is referred to as
\emph{fitted off-shell FSI}.  

The parameters were fitted for each
$(Q^2,W)$ to all measured spectator momenta.  When comparing the results of the
fits to the data
it became clear that the model fits systematically underestimate the data at
the highest measured spectator momentum $p_s=560$ MeV.  This may be a
consequence of the factorization used in this model, which begins to break down
at these momenta (see discussion in Sec. \ref{subsec:approx}).  In subsequent
fits we decided to exclude the highest
spectator momentum as we deem the model not adequate enough to describe the
data in these kinematics.

\begin{figure}[htb]
\begin{center}
  \includegraphics[width=\textwidth]{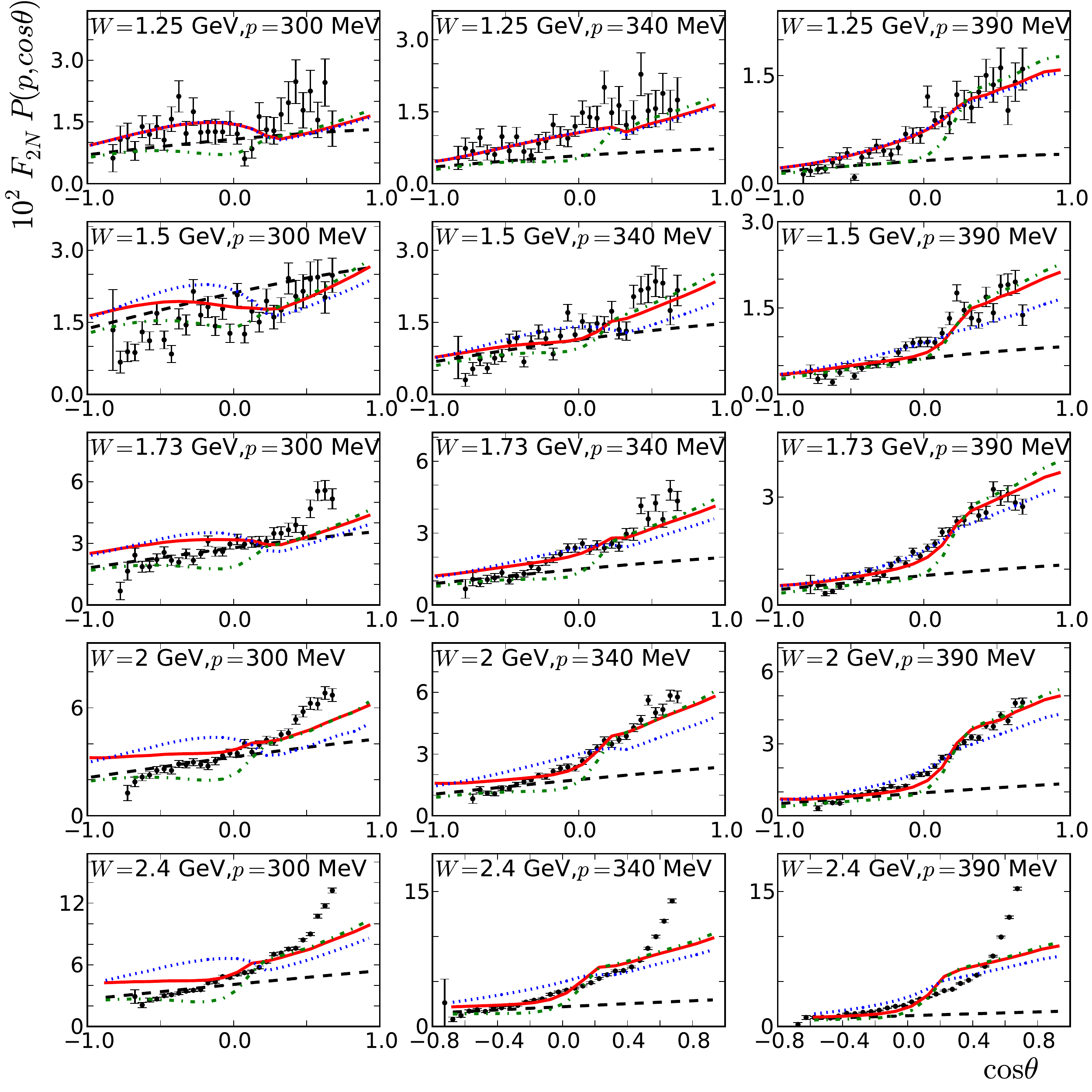}
\caption{(Color online) Comparison between the \emph{Deeps} data
\cite{Klimenko:2005zz} and model calculations at $Q^2=1.8~\text{GeV}^2$ at
measured values of invariant mass $W$ and spectator momenta $p$ 
($\equiv p_s$ in the text) of 300, 340 and 390 MeV. The
dashed black curve is a
plane-wave calculation, the other include final-state interactions.  The
effective total cross section and slope parameter in the final-state
interaction amplitude are fitted parameters for each $W$, the
real part is fixed at $\epsilon=-0.5$.  The dot-dashed green curve only
considers
on-shell
rescattering, the dotted blue curve has an off-shell rescattering amplitude
equal to
the on-shell one and the full red curve uses the off-shell
parameterization of Eq.~(\ref{eq:offrescatter}).}
\label{fig:fit1Q18_low}       
\end{center}
\end{figure}

\begin{figure}[htb]
\begin{center}
  \includegraphics[width=0.7\textwidth]{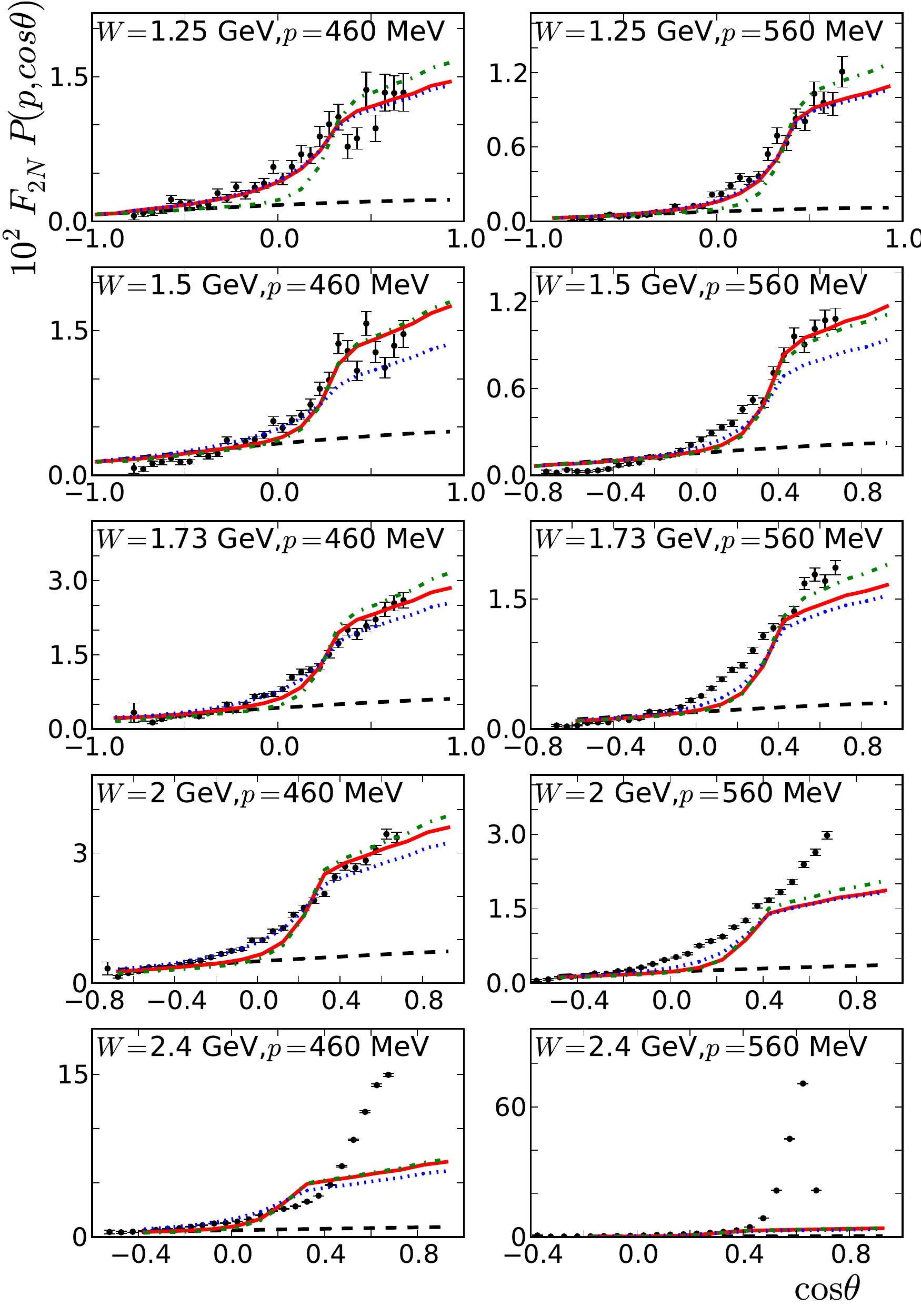}
\caption{(Color online) Comparison between the \emph{Deeps} data
\cite{Klimenko:2005zz} and model calculations at $Q^2=1.8~\text{GeV}^2$ at
measured values of invariant mass $W$ and spectator momenta $p$ 
($\equiv p_s$ in the text) of 460 and 560 MeV. Graphs as in Fig.
\ref{fig:fit1Q18_low}.}
\label{fig:fit1Q18_hi}       
\end{center}
\end{figure}

\begin{figure}[htb]
\begin{center}
  \includegraphics[width=\textwidth]{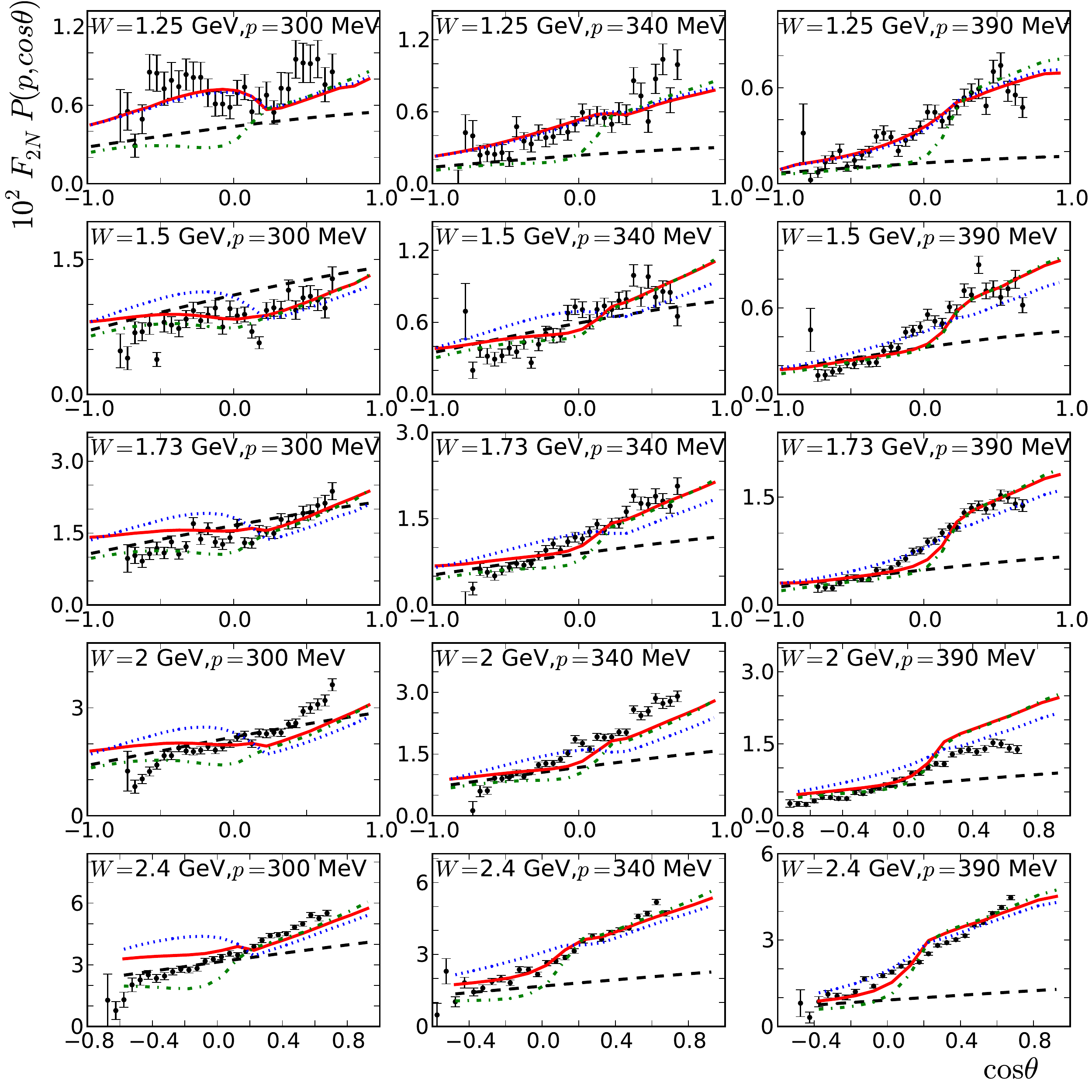}
\caption{(Color online) Comparison between the \emph{Deeps} data
\cite{Klimenko:2005zz} and model calculations at $Q^2=2.8~\text{GeV}^2$ at
measured values of invariant mass $W$ and spectator momenta $p$
($\equiv p_s$ in the text) of 300, 340 and 390 MeV.  Graphs as in Fig.
\ref{fig:fit1Q18_low}.}
\label{fig:fit1Q28_low}       
\end{center}
\end{figure}

\begin{figure}[htb]
\begin{center}
  \includegraphics[width=0.7\textwidth]{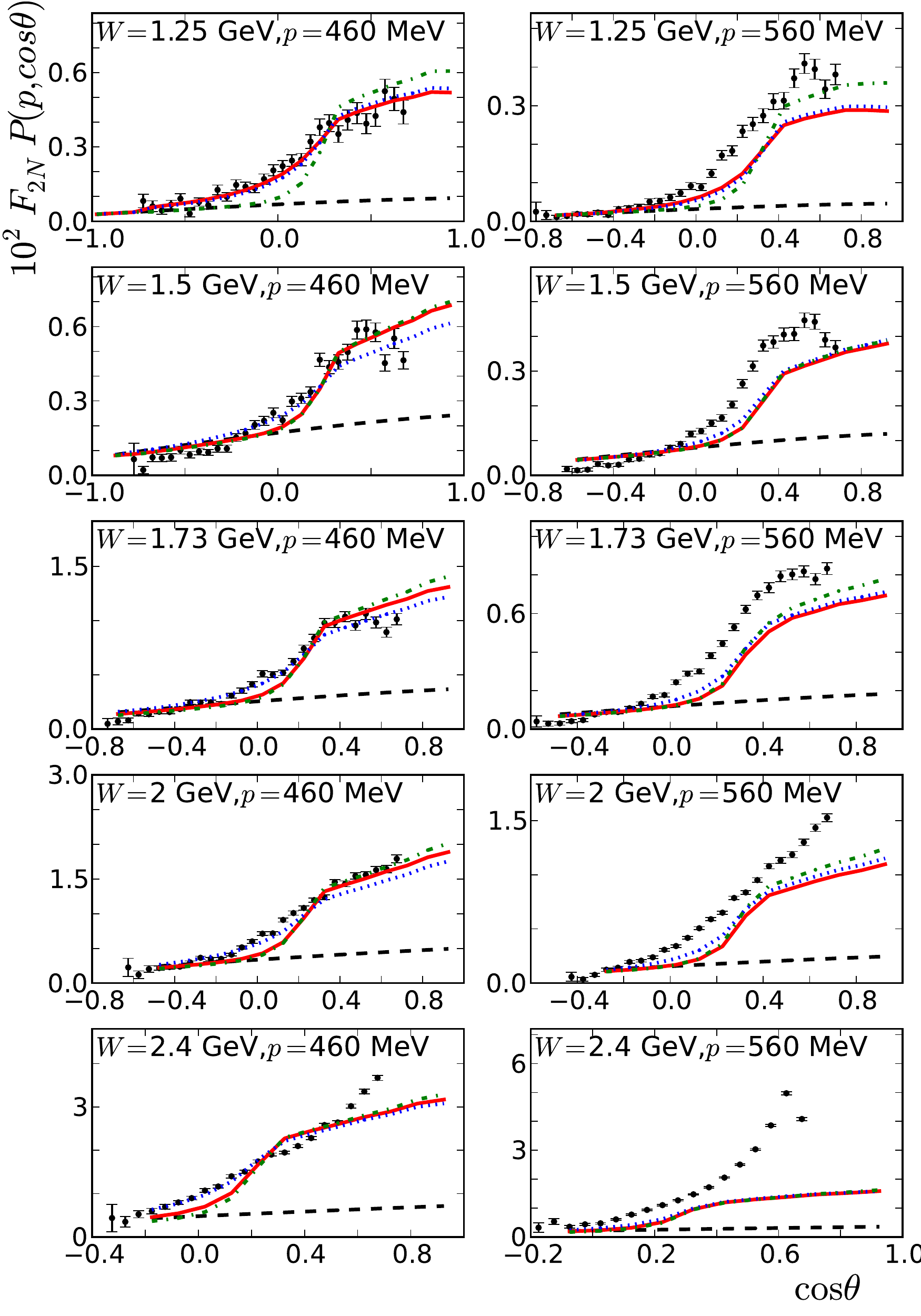}
\caption{(Color online) Comparison between the \emph{Deeps} data
\cite{Klimenko:2005zz} and model calculations at $Q^2=2.8~\text{GeV}^2$ at
measured values of invariant mass $W$ and spectator momenta $p$
($\equiv p_s$ in the text) of 300, 340 and 390 MeV.  Graphs as in Fig.
\ref{fig:fit1Q18_low}.}
\label{fig:fit1Q28_hi}       
\end{center}
\end{figure}

Figures \ref{fig:fit1Q18_low} to \ref{fig:fit1Q28_hi} show the results of these
fits
in which both the effective cross section $\sigma$ and slope factor $B$ were
free
parameters. The plane-wave calculations generally show little dependence on
the spectator angle, in clear disagreement with the change seen in the data. 
The
calculations including
FSI manage fairly well to describe the data over the covered kinematics.  When
comparing the three off-shell descriptions, we see that differences between
the three become smaller with higher spectator momentum.  This indicates the
diminished importance of the off-shell part of the rescattering amplitude in
these kinematics.

At
the lowest missing momentum of $p_s=300$ MeV, there is an oscillating structure
in the data which disappears for high $W$ but is still present in the
calculations.  When comparing the three calculations including FSI, we see that
there is a large difference between them in the backward angles.  There, the
\emph{no off-shell FSI} calculation is smaller than the plane-wave calculations
while the \emph{maximal off-shell FSI} calculation becomes significantly
bigger. The \emph{fitted off-shell FSI} calculations sits somewhere in between
and tends to
agree more with the \emph{maximal off-shell FSI} calculation at low $W$
and with the
\emph{no off-shell} one at high $W$.   At this value of spectator momentum, the
plane-wave and
final-state interaction amplitudes are of comparable magnitudes\footnote{This
situation is 
similar to the quasielastic $d(e,e'N)N$ reaction (see
e.g.\cite{Sargsian:2009hf}).}.   
This makes the final result quite sensitive to small variations in the FSI
amplitude and its
off-shell description, thus providing some way of explaining the larger
discrepancy between data and different calculations as compared to higher $p_s$
values.  

At higher spectator momenta, the \emph{no off-shell
FSI} calculations more or less exhibit three regimes.  At
backward angles they almost coincide with the plain-wave calculations.  Around
90 degrees they show a steep rise, which flattens out at the forward angles. 
This agrees with the intuitive picture of final-state rescattering.  The
\emph{maximal off-shell FSI} calculations on the other hand have a more
constant
slope for the whole of the spectator momentum range.  The calculations with a
\emph{fitted off-shell FSI} description show the best agreement with the data,
which is to be expected as they have an extra free parameter.  Over the whole of
the
kinematics they generally agree more with the \emph{no off-shell
FSI} 
calculations than the \emph{maximal off-shell FSI} ones, pointing at a largely
suppressed
off-shell amplitude.  At the highest
measured spectator momentum the FSI curves systematically underestimate the
data, pointing in the direction of a breakdown of the factorization used in
this model.  

\begin{figure}[htb]
\begin{center}
  \includegraphics[width=\textwidth]{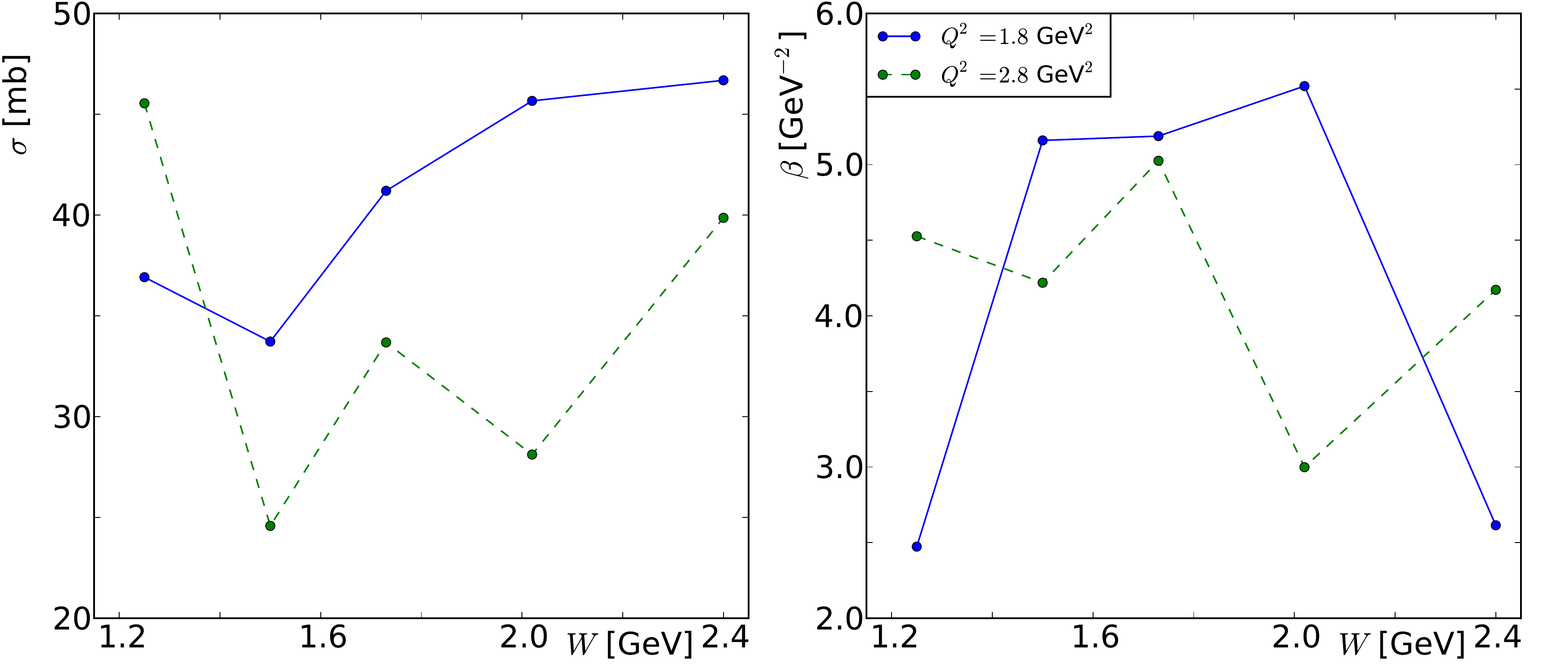}
\caption{(Color online) The fitted values of effective cross section $\sigma$
and slope factor $B$ for the \emph{no off-shell FSI} calculations
used in Figs.~\ref{fig:fit1Q18_low} to \ref{fig:fit1Q28_hi} as a function of the
invariant mass $W$.  Full blue curve is for $Q^2=1.8~\text{GeV}^2$,
the dashed green curve for $Q^2=2.8~\text{GeV}^2$.}
\label{fig:fit2paramon}       
\end{center}
\end{figure}

\begin{figure}[htb]
\begin{center}
  \includegraphics[width=\textwidth]{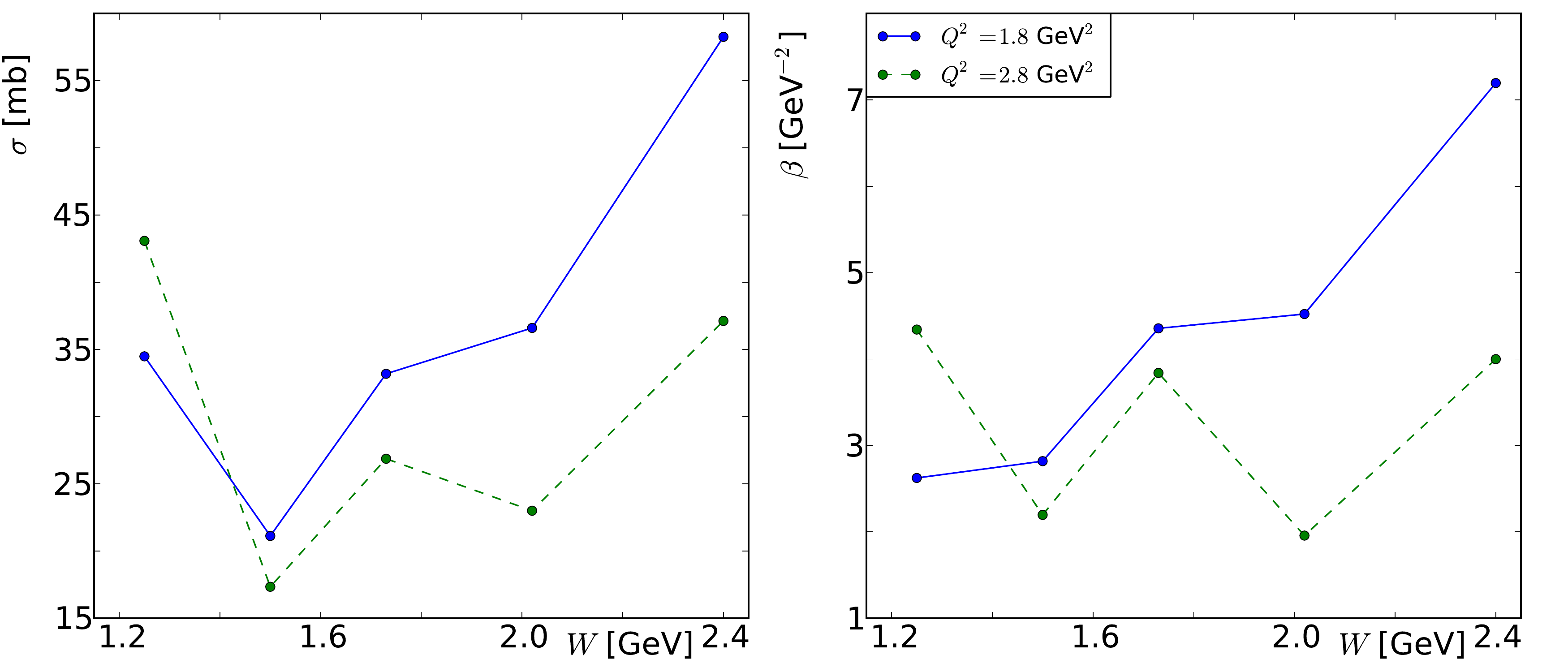}
\caption{(Color online) The fitted values of effective cross section $\sigma$
and slope factor $B$ for the \emph{maximal off-shell} FSI calculations
used in Figs.~\ref{fig:fit1Q18_low} to \ref{fig:fit1Q28_hi} as a function of the
invariant mass $W$.  Full blue curve is for $Q^2=1.8~\text{GeV}^2$,
the dashed green curve for $Q^2=2.8~\text{GeV}^2$.}
\label{fig:fit2param}       
\end{center}
\end{figure}

The final question we addressed in the above described fitting procedure is
whether 
the data indicate on $Q^2$ and $W$ dependence of the parameters of $XN$
rescattering. 

Figures~\ref{fig:fit2paramon} and \ref{fig:fit2param} show the values of the
fitted parameters $\sigma_{\text{tot}}$ and $B$ used in respectively the
\emph{no off-shell}
and
the \emph{maximal off-shell} FSI calculations.  
At $W\approx 1.2$ GeV (corresponding to the production of a
$\Delta$), we get a $\sigma_{\text{tot}}$ around 40~mb.  For the higher
invariant masses, the cross section drops to around 20-25 mb and rises with
increasing
$W$, consistent with the production  of more hadronic constituents in the
intermediate
state of the DIS reaction. The cross section doesn't
flatten out at the highest $W$, showing that hadronization occurs before the
rescattering in these kinematics.  With increasing $Q^2$, the
value of the $(XN)$ cross section parameter also becomes consistently smaller in
this region, indicating reduced final-state interactions.  This could be a sign
of
an onset of a color transparency effect, in which with increasing $Q^2$ the
hadronic state is produced in a state with smaller transverse size, subject to
reduced QCD interactions with the medium.  The values for the slope parameter
$B$ are also largely correlated with those of $\sigma_{\text{tot}}$ with a
smaller slope parameter at higher $Q^2$ and larger $B$ for higher $W$,
although we
also see some clear deviations from this picture (e.g. at $W=2.4$ GeV in the
\emph{no
off-shell} FSI fit).   Overall our fitting procedure indicates that
the availability
of 
more $Q^2$ and $W$ data points may allow to gain important insight about the
$Q^2$ and $W$ dependence of the total cross section of $NX$ scattering.

\section{Conclusion}
\label{sec:conclusion}
Based on the virtual nucleon approximation framework, we developed a model to
describe semi-inclusive deep inelastic scattering of the deuteron.  To
describe the final-state interaction of the spectator nucleon with the produced
hadronic state $X$, the general features of diffractive soft rescattering were
used, without specifying the structure or space-time evolution of $X$.  The
generalized eikonal approximation was used to calculate the scattering
amplitudes based
on effective Feynman diagram rules.  A factorized approach was used to split
the cross section into a part describing the virtual photon interaction with
the off-shell neutron and a distorted spectral function containing the
final-state interactions.  

The model calculations were compared to data taken in the \emph{Deeps}
experiment at
Jefferson Lab.  We first compared our calculation with the data for typical 
kinematics of the experiment with characteristic parameters for final state
interactions.
This comparison indicates a good agreement with the data most importantly 
describing correctly the rise of FSI in forward direction.  This result is 
opposite to what observed in quasi-elastic kinematics.

To gain insight on the $Q^2$ and $W$ evolution of the FSI  
further calculations  were done in which two 
free parameters (effective cross section
$\sigma$ and slope factor $B$) in the rescattering amplitude and three different
off-shell rescattering prescriptions were considered.  
Results were fitted for each $(Q^2,W)$ to the available data.  The fitted
off-shell
rescattering parameterizations yielded results similar to the
calculations with only an on-shell rescattering amplitude included over a wide
range of
the kinematics, giving evidence for
a largely suppressed off-shell rescattering.  
The resulting calculations showed reasonable agreement between the data and the
calculations including final-state interactions.  There were some discrepancies
at the highest spectator momentum, which may be caused by the breakdown of the
factorization used in the model.  At the lowest $p_s=300$ MeV there is also an
oscillating structure in the calculations which isn't exactly present in the
data at higher $W$. The
calculations in this case proved to be very sensitive to the size of the
off-shell amplitude.

When inspecting the values of the parameter fits, three features
emerge:  i)  The effective cross section rises with increasing $W$, consistent
with the creation of more hadronic constituents taking part in the
rescattering. 
ii) There is no evidence for a
plateau at the highest measured $W$ values, indicating that the hadronic state
has hadronized before rescattering takes place.  iii) We obtain lower values
for $\sigma_{\text{tot}}$ for the higher $Q^2$ value, which could be
interpreted as a sign
of emerging color transparency.  However, more data at higher $Q^2$ are needed
to make more definitive statements.

\subsection*{Acknowledgments}

We are thankful to Mark Strikman and Sebastian Kuhn for numerous discussions 
and useful comments in due course of our study.
This work is supported by the Research Foundation Flanders as well as by the  
U.S. Department of Energy Grant 
under Contract DE-FG02-01ER41172.

 \bibliography{../bibtexall.bib}

\begin{thebibliography}{36}
\expandafter\ifx\csname natexlab\endcsname\relax\def\natexlab#1{#1}\fi
\expandafter\ifx\csname bibnamefont\endcsname\relax
  \def\bibnamefont#1{#1}\fi
\expandafter\ifx\csname bibfnamefont\endcsname\relax
  \def\bibfnamefont#1{#1}\fi
\expandafter\ifx\csname citenamefont\endcsname\relax
  \def\citenamefont#1{#1}\fi
\expandafter\ifx\csname url\endcsname\relax
  \def\url#1{\texttt{#1}}\fi
\expandafter\ifx\csname urlprefix\endcsname\relax\def\urlprefix{URL }\fi
\providecommand{\bibinfo}[2]{#2}
\providecommand{\eprint}[2][]{\url{#2}}

\bibitem[{\citenamefont{Melnitchouk et~al.}(1997)\citenamefont{Melnitchouk,
  Sargsian, and Strikman}}]{Melnitchouk:1996vp}
\bibinfo{author}{\bibfnamefont{W.}~\bibnamefont{Melnitchouk}},
  \bibinfo{author}{\bibfnamefont{M.}~\bibnamefont{Sargsian}}, \bibnamefont{and}
  \bibinfo{author}{\bibfnamefont{M.~I.} \bibnamefont{Strikman}},
  \bibinfo{journal}{Z. Phys.} \textbf{\bibinfo{volume}{A359}},
  \bibinfo{pages}{99} (\bibinfo{year}{1997}), \eprint{nucl-th/9609048}.

\bibitem[{\citenamefont{Carlson and Lassila}(1995)}]{Carlson:1994ga}
\bibinfo{author}{\bibfnamefont{C.~E.} \bibnamefont{Carlson}} \bibnamefont{and}
  \bibinfo{author}{\bibfnamefont{K.~E.} \bibnamefont{Lassila}},
  \bibinfo{journal}{Phys. Rev.} \textbf{\bibinfo{volume}{C51}},
  \bibinfo{pages}{364} (\bibinfo{year}{1995}), \eprint{hep-ph/9401307}.

\bibitem[{\citenamefont{Carlson et~al.}(2001)\citenamefont{Carlson, Hanlon, and
  Lassila}}]{Carlson:1999uk}
\bibinfo{author}{\bibfnamefont{C.~E.} \bibnamefont{Carlson}},
  \bibinfo{author}{\bibfnamefont{J.}~\bibnamefont{Hanlon}}, \bibnamefont{and}
  \bibinfo{author}{\bibfnamefont{K.~E.} \bibnamefont{Lassila}},
  \bibinfo{journal}{Phys. Rev.} \textbf{\bibinfo{volume}{D63}},
  \bibinfo{pages}{117301} (\bibinfo{year}{2001}), \eprint{hep-ph/9902281}.

\bibitem[{\citenamefont{Klimenko et~al.}(2006)}]{Klimenko:2005zz}
\bibinfo{author}{\bibfnamefont{A.~V.} \bibnamefont{Klimenko}}
  \bibnamefont{et~al.} (\bibinfo{collaboration}{CLAS}), \bibinfo{journal}{Phys.
  Rev.} \textbf{\bibinfo{volume}{C73}}, \bibinfo{pages}{035212}
  (\bibinfo{year}{2006}), \eprint{nucl-ex/0510032}.

\bibitem[{\citenamefont{Fenker et~al.}(2003)\citenamefont{Fenker, Keppel, Kuhn,
  and Melnitchouk}}]{Bonus:2003}
\bibinfo{author}{\bibfnamefont{H.}~\bibnamefont{Fenker}},
  \bibinfo{author}{\bibfnamefont{C.}~\bibnamefont{Keppel}},
  \bibinfo{author}{\bibfnamefont{S.}~\bibnamefont{Kuhn}}, \bibnamefont{and}
  \bibinfo{author}{\bibfnamefont{W.}~\bibnamefont{Melnitchouk}}
  (\bibinfo{year}{2003}), \bibinfo{note}{jLAB-PR-03-012},
  \urlprefix\url{http://jlab.org/exp_prog/CEBAF_EXP/E03012.html}.

\bibitem[{\citenamefont{Simula}(1996)}]{Simula:1996xk}
\bibinfo{author}{\bibfnamefont{S.}~\bibnamefont{Simula}},
  \bibinfo{journal}{Phys. Lett.} \textbf{\bibinfo{volume}{B387}},
  \bibinfo{pages}{245} (\bibinfo{year}{1996}), \eprint{nucl-th/9605024}.

\bibitem[{\citenamefont{Sargsian and Strikman}(2006)}]{Sargsian:2005rm}
\bibinfo{author}{\bibfnamefont{M.}~\bibnamefont{Sargsian}} \bibnamefont{and}
  \bibinfo{author}{\bibfnamefont{M.}~\bibnamefont{Strikman}},
  \bibinfo{journal}{Phys. Lett.} \textbf{\bibinfo{volume}{B639}},
  \bibinfo{pages}{223} (\bibinfo{year}{2006}), \eprint{hep-ph/0511054}.

\bibitem[{\citenamefont{Ciofi~degli Atti et~al.}(1999)\citenamefont{Ciofi~degli
  Atti, Kaptari, and Scopetta}}]{CiofidegliAtti:1999kp}
\bibinfo{author}{\bibfnamefont{C.}~\bibnamefont{Ciofi~degli Atti}},
  \bibinfo{author}{\bibfnamefont{L.~P.} \bibnamefont{Kaptari}},
  \bibnamefont{and} \bibinfo{author}{\bibfnamefont{S.}~\bibnamefont{Scopetta}},
  \bibinfo{journal}{Eur. Phys. J.} \textbf{\bibinfo{volume}{A5}},
  \bibinfo{pages}{191} (\bibinfo{year}{1999}), \eprint{hep-ph/9904486}.

\bibitem[{\citenamefont{Ciofi~degli Atti and
  Kopeliovich}(2003)}]{CiofidegliAtti:2002as}
\bibinfo{author}{\bibfnamefont{C.}~\bibnamefont{Ciofi~degli Atti}}
  \bibnamefont{and} \bibinfo{author}{\bibfnamefont{B.~Z.}
  \bibnamefont{Kopeliovich}}, \bibinfo{journal}{Eur. Phys. J.}
  \textbf{\bibinfo{volume}{A17}}, \bibinfo{pages}{133} (\bibinfo{year}{2003}),
  \eprint{nucl-th/0207001}.

\bibitem[{\citenamefont{Ciofi~degli Atti et~al.}(2004)\citenamefont{Ciofi~degli
  Atti, Kaptari, and Kopeliovich}}]{CiofidegliAtti:2003pb}
\bibinfo{author}{\bibfnamefont{C.}~\bibnamefont{Ciofi~degli Atti}},
  \bibinfo{author}{\bibfnamefont{L.~P.} \bibnamefont{Kaptari}},
  \bibnamefont{and} \bibinfo{author}{\bibfnamefont{B.~Z.}
  \bibnamefont{Kopeliovich}}, \bibinfo{journal}{Eur. Phys. J.}
  \textbf{\bibinfo{volume}{A19}}, \bibinfo{pages}{145} (\bibinfo{year}{2004}),
  \eprint{nucl-th/0307052}.

\bibitem[{\citenamefont{Palli et~al.}(2009)\citenamefont{Palli, Ciofi~degli
  Atti, Kaptari, Mezzetti, and Alvioli}}]{Palli:2009it}
\bibinfo{author}{\bibfnamefont{V.}~\bibnamefont{Palli}},
  \bibinfo{author}{\bibfnamefont{C.}~\bibnamefont{Ciofi~degli Atti}},
  \bibinfo{author}{\bibfnamefont{L.~P.} \bibnamefont{Kaptari}},
  \bibinfo{author}{\bibfnamefont{C.~B.} \bibnamefont{Mezzetti}},
  \bibnamefont{and} \bibinfo{author}{\bibfnamefont{M.}~\bibnamefont{Alvioli}},
  \bibinfo{journal}{Phys. Rev.} \textbf{\bibinfo{volume}{C80}},
  \bibinfo{pages}{054610} (\bibinfo{year}{2009}), \eprint{0911.1377}.

\bibitem[{\citenamefont{Ciofi~degli Atti and Kaptari}(2011)}]{Atti:2010yf}
\bibinfo{author}{\bibfnamefont{C.}~\bibnamefont{Ciofi~degli Atti}}
  \bibnamefont{and} \bibinfo{author}{\bibfnamefont{L.~P.}
  \bibnamefont{Kaptari}}, \bibinfo{journal}{Phys. Rev.}
  \textbf{\bibinfo{volume}{C83}}, \bibinfo{pages}{044602}
  (\bibinfo{year}{2011}), \eprint{1011.5960}.

\bibitem[{\citenamefont{Frankfurt et~al.}(1995)\citenamefont{Frankfurt,
  Greenberg, Miller, Sargsian, and Strikman}}]{Frankfurt:1994kt}
\bibinfo{author}{\bibfnamefont{L.~L.} \bibnamefont{Frankfurt}},
  \bibinfo{author}{\bibfnamefont{W.~R.} \bibnamefont{Greenberg}},
  \bibinfo{author}{\bibfnamefont{G.~A.} \bibnamefont{Miller}},
  \bibinfo{author}{\bibfnamefont{M.~M.} \bibnamefont{Sargsian}},
  \bibnamefont{and} \bibinfo{author}{\bibfnamefont{M.~I.}
  \bibnamefont{Strikman}}, \bibinfo{journal}{Z. Phys.}
  \textbf{\bibinfo{volume}{A352}}, \bibinfo{pages}{97} (\bibinfo{year}{1995}),
  \eprint{nucl-th/9501009}.

\bibitem[{\citenamefont{Frankfurt
  et~al.}(1997{\natexlab{a}})\citenamefont{Frankfurt, Sargsian, and
  Strikman}}]{Frankfurt:1996xx}
\bibinfo{author}{\bibfnamefont{L.~L.} \bibnamefont{Frankfurt}},
  \bibinfo{author}{\bibfnamefont{M.~M.} \bibnamefont{Sargsian}},
  \bibnamefont{and} \bibinfo{author}{\bibfnamefont{M.~I.}
  \bibnamefont{Strikman}}, \bibinfo{journal}{Phys. Rev.}
  \textbf{\bibinfo{volume}{C56}}, \bibinfo{pages}{1124}
  (\bibinfo{year}{1997}{\natexlab{a}}), \eprint{nucl-th/9603018}.

\bibitem[{\citenamefont{Sargsian}(2001)}]{Sargsian:2001ax}
\bibinfo{author}{\bibfnamefont{M.~M.} \bibnamefont{Sargsian}},
  \bibinfo{journal}{Int. J. Mod. Phys.} \textbf{\bibinfo{volume}{E10}},
  \bibinfo{pages}{405} (\bibinfo{year}{2001}), \eprint{nucl-th/0110053}.

\bibitem[{\citenamefont{Sargsian et~al.}(2002)\citenamefont{Sargsian, Simula,
  and Strikman}}]{Sargsian:2001gu}
\bibinfo{author}{\bibfnamefont{M.~M.} \bibnamefont{Sargsian}},
  \bibinfo{author}{\bibfnamefont{S.}~\bibnamefont{Simula}}, \bibnamefont{and}
  \bibinfo{author}{\bibfnamefont{M.~I.} \bibnamefont{Strikman}},
  \bibinfo{journal}{Phys. Rev.} \textbf{\bibinfo{volume}{C66}},
  \bibinfo{pages}{024001} (\bibinfo{year}{2002}), \eprint{nucl-th/0105052}.

\bibitem[{\citenamefont{Sargsian}(2010)}]{Sargsian:2009hf}
\bibinfo{author}{\bibfnamefont{M.~M.} \bibnamefont{Sargsian}},
  \bibinfo{journal}{Phys. Rev.} \textbf{\bibinfo{volume}{C82}},
  \bibinfo{pages}{014612} (\bibinfo{year}{2010}), \eprint{0910.2016}.

\bibitem[{\citenamefont{Sargsian et~al.}(2003)}]{Sargsian:2002wc}
\bibinfo{author}{\bibfnamefont{M.~M.} \bibnamefont{Sargsian}}
  \bibnamefont{et~al.}, \bibinfo{journal}{J. Phys.}
  \textbf{\bibinfo{volume}{G29}}, \bibinfo{pages}{R1} (\bibinfo{year}{2003}),
  \eprint{nucl-th/0210025}.

\bibitem[{\citenamefont{Frankfurt and Strikman}(1981)}]{Frankfurt1981215}
\bibinfo{author}{\bibfnamefont{L.~L.} \bibnamefont{Frankfurt}}
  \bibnamefont{and} \bibinfo{author}{\bibfnamefont{M.~I.}
  \bibnamefont{Strikman}}, \bibinfo{journal}{Physics Reports}
  \textbf{\bibinfo{volume}{76}}, \bibinfo{pages}{215 } (\bibinfo{year}{1981}).

\bibitem[{\citenamefont{Frankfurt and Strikman}(1976)}]{Frankfurt:1976gz}
\bibinfo{author}{\bibfnamefont{L.~L.} \bibnamefont{Frankfurt}}
  \bibnamefont{and} \bibinfo{author}{\bibfnamefont{M.~I.}
  \bibnamefont{Strikman}}, \bibinfo{journal}{Phys. Lett.}
  \textbf{\bibinfo{volume}{B64}}, \bibinfo{pages}{433} (\bibinfo{year}{1976}).

\bibitem[{\citenamefont{Frankfurt and Strikman}(1987)}]{Frankfurt1987254}
\bibinfo{author}{\bibfnamefont{L.~L.} \bibnamefont{Frankfurt}}
  \bibnamefont{and} \bibinfo{author}{\bibfnamefont{M.~I.}
  \bibnamefont{Strikman}}, \bibinfo{journal}{Physics Letters B}
  \textbf{\bibinfo{volume}{183}}, \bibinfo{pages}{254 } (\bibinfo{year}{1987}).

\bibitem[{\citenamefont{Landshoff and Polkinghorne}(1978)}]{Landshoff:1977pg}
\bibinfo{author}{\bibfnamefont{P.~V.} \bibnamefont{Landshoff}}
  \bibnamefont{and} \bibinfo{author}{\bibfnamefont{J.~C.}
  \bibnamefont{Polkinghorne}}, \bibinfo{journal}{Phys. Rev.}
  \textbf{\bibinfo{volume}{D18}}, \bibinfo{pages}{153} (\bibinfo{year}{1978}).

\bibitem[{\citenamefont{Farrar et~al.}(1988)\citenamefont{Farrar, Liu,
  Frankfurt, and Strikman}}]{Farrar:1988me}
\bibinfo{author}{\bibfnamefont{G.~R.} \bibnamefont{Farrar}},
  \bibinfo{author}{\bibfnamefont{H.}~\bibnamefont{Liu}},
  \bibinfo{author}{\bibfnamefont{L.~L.} \bibnamefont{Frankfurt}},
  \bibnamefont{and} \bibinfo{author}{\bibfnamefont{M.~I.}
  \bibnamefont{Strikman}}, \bibinfo{journal}{Phys. Rev. Lett.}
  \textbf{\bibinfo{volume}{61}}, \bibinfo{pages}{686} (\bibinfo{year}{1988}).

\bibitem[{\citenamefont{Frankfurt et~al.}(1994)\citenamefont{Frankfurt, Miller,
  and Strikman}}]{Frankfurt:1994hf}
\bibinfo{author}{\bibfnamefont{L.~L.} \bibnamefont{Frankfurt}},
  \bibinfo{author}{\bibfnamefont{G.~A.} \bibnamefont{Miller}},
  \bibnamefont{and} \bibinfo{author}{\bibfnamefont{M.}~\bibnamefont{Strikman}},
  \bibinfo{journal}{Ann. Rev. Nucl. Part. Sci.} \textbf{\bibinfo{volume}{44}},
  \bibinfo{pages}{501} (\bibinfo{year}{1994}), \eprint{hep-ph/9407274}.

\bibitem[{\citenamefont{Frankfurt
  et~al.}(1997{\natexlab{b}})}]{Frankfurt:1997ss}
\bibinfo{author}{\bibfnamefont{L.}~\bibnamefont{Frankfurt}}
  \bibnamefont{et~al.}, \bibinfo{journal}{Nucl. Phys.}
  \textbf{\bibinfo{volume}{A622}}, \bibinfo{pages}{511}
  (\bibinfo{year}{1997}{\natexlab{b}}), \eprint{hep-ph/9703399}.

\bibitem[{\citenamefont{Cosyn et~al.}(2008)\citenamefont{Cosyn, Martinez, and
  Ryckebusch}}]{Cosyn:2007er}
\bibinfo{author}{\bibfnamefont{W.}~\bibnamefont{Cosyn}},
  \bibinfo{author}{\bibfnamefont{M.~C.} \bibnamefont{Martinez}},
  \bibnamefont{and}
  \bibinfo{author}{\bibfnamefont{J.}~\bibnamefont{Ryckebusch}},
  \bibinfo{journal}{Phys. Rev.} \textbf{\bibinfo{volume}{C77}},
  \bibinfo{pages}{034602} (\bibinfo{year}{2008}), \eprint{0710.4837}.

\bibitem[{\citenamefont{Ryckebusch et~al.}(2007)\citenamefont{Ryckebusch,
  Cosyn, Van~Overmeire, and Martinez}}]{Ryckebusch:2007zz}
\bibinfo{author}{\bibfnamefont{J.}~\bibnamefont{Ryckebusch}},
  \bibinfo{author}{\bibfnamefont{W.}~\bibnamefont{Cosyn}},
  \bibinfo{author}{\bibfnamefont{B.}~\bibnamefont{Van~Overmeire}},
  \bibnamefont{and} \bibinfo{author}{\bibfnamefont{C.}~\bibnamefont{Martinez}},
  \bibinfo{journal}{Eur. Phys. J.} \textbf{\bibinfo{volume}{A31}},
  \bibinfo{pages}{585} (\bibinfo{year}{2007}).

\bibitem[{\citenamefont{Gallmeister et~al.}(2011)\citenamefont{Gallmeister,
  Kaskulov, and Mosel}}]{Gallmeister:2010wn}
\bibinfo{author}{\bibfnamefont{K.}~\bibnamefont{Gallmeister}},
  \bibinfo{author}{\bibfnamefont{M.}~\bibnamefont{Kaskulov}}, \bibnamefont{and}
  \bibinfo{author}{\bibfnamefont{U.}~\bibnamefont{Mosel}},
  \bibinfo{journal}{Phys. Rev.} \textbf{\bibinfo{volume}{C83}},
  \bibinfo{pages}{015201} (\bibinfo{year}{2011}), \eprint{1007.1141}.

\bibitem[{\citenamefont{Frankfurt et~al.}(1992)\citenamefont{Frankfurt,
  Greenberg, Miller, and Strikman}}]{Frankfurt:1992zp}
\bibinfo{author}{\bibfnamefont{L.}~\bibnamefont{Frankfurt}},
  \bibinfo{author}{\bibfnamefont{W.~R.} \bibnamefont{Greenberg}},
  \bibinfo{author}{\bibfnamefont{G.~A.} \bibnamefont{Miller}},
  \bibnamefont{and} \bibinfo{author}{\bibfnamefont{M.}~\bibnamefont{Strikman}},
  \bibinfo{journal}{Phys. Rev.} \textbf{\bibinfo{volume}{C46}},
  \bibinfo{pages}{2547} (\bibinfo{year}{1992}), \eprint{nucl-th/9211002}.

\bibitem[{\citenamefont{Jeschonnek}(2001)}]{Jeschonnek:2000nh}
\bibinfo{author}{\bibfnamefont{S.}~\bibnamefont{Jeschonnek}},
  \bibinfo{journal}{Phys. Rev.} \textbf{\bibinfo{volume}{C63}},
  \bibinfo{pages}{034609} (\bibinfo{year}{2001}), \eprint{nucl-th/0009086}.

\bibitem[{\citenamefont{Gribov}(1970)}]{Gribov:1968gs}
\bibinfo{author}{\bibfnamefont{V.~N.} \bibnamefont{Gribov}},
  \bibinfo{journal}{Sov. Phys. JETP} \textbf{\bibinfo{volume}{30}},
  \bibinfo{pages}{709} (\bibinfo{year}{1970}).

\bibitem[{\citenamefont{Bertocchi}(1972)}]{Bertocchi:1972}
\bibinfo{author}{\bibfnamefont{L.}~\bibnamefont{Bertocchi}},
  \bibinfo{journal}{Il Nuovo Cimento A (1971-1996)}
  \textbf{\bibinfo{volume}{11}}, \bibinfo{pages}{45} (\bibinfo{year}{1972}),
  ISSN \bibinfo{issn}{0369-3546}, \bibinfo{note}{10.1007/BF02722777},
  \urlprefix\url{http://dx.doi.org/10.1007/BF02722777}.

\bibitem[{\citenamefont{Machleidt}(2001)}]{Machleidt:2000ge}
\bibinfo{author}{\bibfnamefont{R.}~\bibnamefont{Machleidt}},
  \bibinfo{journal}{Phys. Rev.} \textbf{\bibinfo{volume}{C63}},
  \bibinfo{pages}{024001} (\bibinfo{year}{2001}), \eprint{nucl-th/0006014}.

\bibitem[{\citenamefont{Lacombe et~al.}(1980)}]{Lacombe:1980dr}
\bibinfo{author}{\bibfnamefont{M.}~\bibnamefont{Lacombe}} \bibnamefont{et~al.},
  \bibinfo{journal}{Phys. Rev.} \textbf{\bibinfo{volume}{C21}},
  \bibinfo{pages}{861} (\bibinfo{year}{1980}).

\bibitem[{\citenamefont{Bodek et~al.}(1979)\citenamefont{Bodek, Breidenbach,
  Dubin, Elias, Friedman, Kendall, Poucher, Riordan, Sogard, Coward
  et~al.}}]{PhysRevD.20.1471}
\bibinfo{author}{\bibfnamefont{A.}~\bibnamefont{Bodek}},
  \bibinfo{author}{\bibfnamefont{M.}~\bibnamefont{Breidenbach}},
  \bibinfo{author}{\bibfnamefont{D.~L.} \bibnamefont{Dubin}},
  \bibinfo{author}{\bibfnamefont{J.~E.} \bibnamefont{Elias}},
  \bibinfo{author}{\bibfnamefont{J.~I.} \bibnamefont{Friedman}},
  \bibinfo{author}{\bibfnamefont{H.~W.} \bibnamefont{Kendall}},
  \bibinfo{author}{\bibfnamefont{J.~S.} \bibnamefont{Poucher}},
  \bibinfo{author}{\bibfnamefont{E.~M.} \bibnamefont{Riordan}},
  \bibinfo{author}{\bibfnamefont{M.~R.} \bibnamefont{Sogard}},
  \bibinfo{author}{\bibfnamefont{D.~H.} \bibnamefont{Coward}},
  \bibnamefont{et~al.}, \bibinfo{journal}{Phys. Rev. D}
  \textbf{\bibinfo{volume}{20}}, \bibinfo{pages}{1471} (\bibinfo{year}{1979}).

\bibitem[{\citenamefont{Frankfurt et~al.}(2008)\citenamefont{Frankfurt,
  Sargsian, and Strikman}}]{Frankfurt:2008zv}
\bibinfo{author}{\bibfnamefont{L.}~\bibnamefont{Frankfurt}},
  \bibinfo{author}{\bibfnamefont{M.}~\bibnamefont{Sargsian}}, \bibnamefont{and}
  \bibinfo{author}{\bibfnamefont{M.}~\bibnamefont{Strikman}},
  \bibinfo{journal}{Int. J. Mod. Phys.} \textbf{\bibinfo{volume}{A23}},
  \bibinfo{pages}{2991} (\bibinfo{year}{2008}), \eprint{0806.4412}.

\end{thebibliography}

\end{document}